\documentclass[a4paper,11pt]{article}
\pdfoutput=1 

\usepackage{jcappub} 

\usepackage[T1]{fontenc} 
\usepackage[dvipsnames]{xcolor}
\usepackage{amsmath,amssymb,amsfonts,mathtools,bm}

\newcommand{\dd}{\mathrm{d}}

\title{\boldmath Primordial spectra from modified Bekenstein--Hawking entropy law}

\author[a,b]{Marco de Cesare,}
\author[c,b]{Giulia Gubitosi}
\author[d,1]{and Varun Kushwaha%
\note{Corresponding author.}}

\affiliation[a]{\emph{Scuola Superiore Meridionale,}\\
\emph{Largo San Marcellino 10, 80138 Napoli, Italy}}

\affiliation[b]{\emph{INFN, Sezione di Napoli,}\\
\emph{Monte S. Angelo, Via Cintia, 80126 Napoli, Italy}}

\affiliation[c]{\emph{Dipartimento di Fisica ``Ettore Pancini'',
Università di Napoli ``Federico II'',}\\
\emph{Monte S. Angelo, Via Cintia, 80126 Napoli, Italy}}

\affiliation[d]{\emph{Faculty of Physics,
Ludwig-Maximilians-Universität,}\\
\emph{Scheinerstraße 1, 81677 Munich, Germany}}

\emailAdd{marco.decesare@na.infn.it}
\emailAdd{giulia.gubitosi@unina.it}
\emailAdd{varun.kushwaha@lmu.de}

\abstract{
Many approaches to quantum gravity predict logarithmic corrections to the Bekenstein--Hawking entropy. Within the spacetime-thermodynamic description of gravity, such corrections lead to modified gravitational field equations. We study their effect on primordial perturbations during standard single-field slow-roll inflation. We derive the evolution equation for the comoving curvature perturbation and show that it retains the standard Mukhanov--Sasaki form, with the quantum-gravity correction entering through the time-dependent effective frequency. We compute the scalar and tensor primordial spectra at next-to-next-to-next-to-leading order ($\mathrm{N}^3\mathrm{LO}$) in the Hubble-flow parameters, keeping the leading contribution from the logarithmic correction. The scalar spectrum remains nearly scale invariant, but the quantum-gravity correction shifts its tilt and runnings. Free tensor modes still propagate as in general relativity, although on the quantum-gravity-corrected background. This different response of the two sectors shifts both the tensor-to-scalar ratio and the single-field consistency relation, which provide the starting point for tracing the model's signatures through the post-inflationary evolution and into the CMB.}

\begin{document}
\maketitle
\raggedbottom

\section{Introduction}
\label{sec:intro}

Logarithmic corrections to the Bekenstein--Hawking entropy appear in many approaches to quantum gravity, including loop quantum gravity, AdS/CFT correspondence, string theory, and semiclassical gravity \cite{Kaul_2000,Carlip_2000,Meissner_2004,Sen_2013,Banerjee_2011}. They also arise in calculations of entanglement entropy across horizons \cite{Bombelli:1986rw,Srednicki_1993,Solodukhin:2011gn}. Although their microscopic origin differs from one framework to another, they suggest that the area law may only be the leading term in the entropy of a gravitational horizon.

Jacobson's framework of spacetime thermodynamics provides a direct way to ask how such subleading contributions to the horizon entropy
modify gravitational dynamics. The Einstein equations of general relativity follow by applying the Clausius relation to local causal horizons together with the standard
Bekenstein--Hawking area--entropy relation \cite{Jacobson_1995,Jacobson_2003,Eling_2006,Padmanabhan_2005}. Replacing this relation with its logarithmically corrected form leads instead to the ``quantum-phenomenological gravitational field equations'' derived in Refs.~\cite{Alonso_Serrano_2020,Alonso_Serrano_2023}. The logarithmic term in the horizon entropy introduces higher-curvature corrections to the field equations, controlled by a dimensionless parameter $\alpha$\,.\footnote{The higher-curvature coupling in Refs.~\cite{Alonso_Serrano_2020,Alonso_Serrano_2023} is written as $D\ell_{\rm P}^{2}$\,, with $\alpha=D/(8\pi)$ in our conventions. For example, the loop-quantum-gravity logarithmic correction obtained in Ref.~\cite{Kaul_2000} corresponds to $\alpha\simeq-1.1\times10^{-3}$\,. We nevertheless treat $\alpha$ as a free phenomenological parameter, since its magnitude and sign depend on the microscopic origin of the entropy correction.} Setting $\alpha=0$ removes these terms and recovers the trace-free Einstein equations of unimodular gravity \cite{Alonso_Serrano_2025}.

The cosmological consequences of these equations were studied in Ref.~\cite{decesare2023cosmologicalevolutionmodifiedbekenstein}. On a spatially flat FLRW background, they admit self-accelerating solutions at both early and late times. The early-time branch, however, does not remove the need for conventional inflation. It neither replaces the initial singularity with a bounce nor produces slow-roll expansion unless the dominant matter already has an equation of state close to $w=-1$\,. Therefore, the deviations from general relativity encoded in the model are more naturally viewed as modifying the standard inflationary dynamics, which is still driven by a separate matter sector. 

In this work, we take the matter sector to be a minimally coupled scalar field with an arbitrary potential, while the background spacetime and its perturbations obey the modified gravitational equations of Refs.~\cite{Alonso_Serrano_2020,Alonso_Serrano_2023}. Inflation provides a sensitive regime in which to study these terms because their relative importance grows with the curvature scale. They can therefore affect both the inflationary background and the scalar and tensor perturbations evolving on it.

For scalar perturbations, we derive a closed evolution equation for the comoving curvature perturbation and show that it retains the standard Mukhanov--Sasaki form. The modification is encoded in an effective background function $z_{\rm eff}$\,, which changes the term $z_{\rm eff}''/z_{\rm eff}$ in the scalar mode equation. Therefore, the propagation speed of the scalar modes remains equal to unity, while the time-dependent part of their mode frequency is modified. This form allows us to use the the Green's function method of Refs.~\cite{Stewart_2001,Auclair_2022} to compute the scalar spectrum through next-to-next-to-next-to-leading order, $\mathrm{N}^3\mathrm{LO}$\,. Carrying the calculation to third order is necessary because the first explicit correction to the scalar running appears only at that order. In the tensor sector, free modes obey the same propagation equation as in general relativity in the absence of anisotropic stress, although on the modified inflationary background.

The Green's function method has previously been applied to models with non-canonical kinetic terms, time-dependent sound speed, and more general effective descriptions of inflation \cite{WeiCaiWang_2004,Hu_2011,MotohashiHu_2017,Auclair_2022,Bianchi_2024}. The difference here is that the mode equation follows directly from modified gravitational field equations that are not presently known to arise from a covariant action. The evolution equations therefore do not by themselves determine the absolute normalisation of the spectra. We specify the normalisation choices when presenting the scalar and tensor spectra and return to their implications in the discussion.

Our main results are the scalar and tensor primordial spectra through $\mathrm{N}^3\mathrm{LO}$\,. We define $\delta\coloneqq\alpha\kappa H^2$\,, where $\kappa=8\pi G$ and $H$ is the Hubble rate. This parameter measures the strength of the correction at the inflationary curvature scale, while its explicit contribution to the scalar mode equation is additionally suppressed during slow roll and vanishes in the exact de~Sitter limit. For smoothly evolving slow-roll backgrounds, the scalar spectrum retains its usual expansion in powers of $\ln(k/k_\diamond)$\,, where $k_\diamond=0.05,{\rm Mpc}^{-1}$ is the pivot scale. The correction changes the amplitude, tilt, and runnings without introducing a new spectral shape. It first contributes explicitly to the tilt at second slow-roll order and to the running at third order. The modified background affects both the scalar and tensor spectra, but only the scalar mode equation contains an additional explicit correction to its time-dependent frequency. The resulting difference between the two sectors modifies both the tensor-to-scalar ratio and the single-field consistency relation.

The paper is organised as follows. In Section~\ref{sec:review}, we review the modified field equations, their FLRW background dynamics, and the linear scalar and tensor perturbation equations. In Section~\ref{sec:modified_ms_equation}, we specialise to single-field inflation and derive the effective Mukhanov--Sasaki equation. In Section~\ref{sec:N3LO_power_spectra}, we calculate the scalar and tensor spectra and obtain the scalar observables, the tensor-to-scalar ratio, and the modified consistency relation through $\mathrm{N}^3\mathrm{LO}$\,. In Section~\ref{sec:discussion_outlook}, we discuss the interpretation and limitations of the calculation and outline the next steps. Appendix~\ref{app:GSR_details} contains the details of the Green's function expansion, the transport to the pivot scale, and the normalisation of the scalar spectrum.

Throughout the paper, we use units in which $c=\hbar=1$ and write $\kappa=8\pi G$\,. The metric signature is $(-,+,+,+)$\,. An overdot and a prime denote derivatives with respect to cosmic and conformal time, respectively. Complex conjugation is denoted by a superscript $*$\,, and evaluation at the pivot time, defined by $-k_\diamond\eta_\diamond=1$\,, is indicated by a subscript $\diamond$\,.

\section{Quantum-gravity corrections to cosmological dynamics}
\label{sec:review}
We begin by summarising the equations that govern the cosmological background and its linear perturbations. The thermodynamic construction of Ref.~\cite{Alonso_Serrano_2020} leads to the trace-free field equations
\begin{align}
    \mathcal{E}_{\mu}{}^\nu  \coloneqq S_{\mu}{}^\nu  - \alpha\kappa S_{\mu\rho}S^{\rho\nu} + \frac{\alpha\kappa}{4}\left(R_{\rho\sigma}R^{\rho\sigma} -\frac{R^2}{4} \right)\delta_{\mu}{}^\nu = \kappa\left( T_{\mu}{}^\nu-\frac{T}{4}\delta_{\mu}{}^\nu \right)~,
    \label{eq:entropy_corrected_field_equations}
\end{align}
where $S_{\mu\nu}    \coloneqq    R_{\mu\nu}-\frac{R}{4}g_{\mu\nu}$ is the traceless part of the Ricci tensor. Here $T_{\mu\nu}$ is the matter stress-energy tensor, and $T=T_\mu{}^\mu$ is its trace. The parameter $\alpha$ measures the strength of the higher-curvature
terms induced by the logarithmic contribution to the horizon entropy. In a microscopic theory, both the magnitude and the sign of $\alpha$ would be fixed by the coefficient of the logarithmic term. In this work we adopt a phenomenological viewpoint and treat $\alpha$ as a free parameter.

For $\alpha = 0$\,, Eq.~\eqref{eq:entropy_corrected_field_equations} reduces to the trace-free Einstein equations of unimodular gravity \cite{Alonso_Serrano_2025}. In that limit, stress-energy conservation allows one to recover the usual Einstein equations, with the cosmological constant appearing as an integration constant. For $\alpha\neq0$\,, the curvature-squared terms modify the trace-free equations themselves, so this reconstruction no longer goes through in the same way.

Moreover, Eq.~\eqref{eq:entropy_corrected_field_equations} is not presently known to arise from a covariant action. We therefore impose matter conservation independently,
\begin{align}
\nabla_\nu T_{\mu}{}^\nu = 0~.
\label{eq:stress_energy_conservation}
\end{align}
Together, Eqs.~\eqref{eq:entropy_corrected_field_equations} and \eqref{eq:stress_energy_conservation} define the effective gravitational dynamics used throughout this paper.

For the background cosmology reviewed below, matter may be modelled as a perfect fluid,
\begin{align}
    T_{\mu\nu}=(\rho+p)u_\mu u_\nu+p g_{\mu\nu}~,
    \label{eq:perfect_fluid_stress_tensor}
\end{align}
where $u^\mu$ is the fluid four-velocity, $\rho$ is the energy density, and $p$ is the pressure. In the inflationary application considered in later sections, this stress-energy tensor will be specialised to that of a minimally coupled scalar field.

\subsection{Background dynamics}
\label{sec:background_cosmology}

The cosmological background and perturbation theory associated with  Eq.~\eqref{eq:entropy_corrected_field_equations} were analysed in detail in  Ref.~\cite{decesare2023cosmologicalevolutionmodifiedbekenstein}. Here we recall the equations and results needed for our calculations. For a spatially flat Friedmann--Lemaître--Robertson--Walker (FLRW) spacetime in conformal time,
\begin{align}
    \dd s^2  =  a^2(\eta) \left(-\dd\eta^2+ \delta_{ij}\dd x^i\dd x^j\right)~,
    \label{eq:flrw_metric}
\end{align}
the conformal Hubble parameter is $\mathcal H =\frac{a'}{a}$\,, where a prime denotes differentiation with respect to conformal time $\eta$\,, related to cosmic time by $\dd t=a\,\dd\eta$\,.

Substituting Eq.~\eqref{eq:flrw_metric} and the perfect-fluid stress tensor \eqref{eq:perfect_fluid_stress_tensor} into Eq.~\eqref{eq:entropy_corrected_field_equations}, one obtains the modified Raychaudhuri equation
\begin{align}
    \mathcal{H}' - \mathcal{H}^2-\frac{\alpha\kappa}{a^2}\left(\mathcal{H}' - \mathcal{H}^2\right)^2=-\frac{\kappa}{2}a^2(\bar{\rho}+\bar p)~,
    \label{eq:modified_raychaudhuri}
\end{align}
where an overbar denotes a background quantity. In the limit $\alpha\to0$\,, Eq.~\eqref{eq:modified_raychaudhuri} reduces to the standard trace-free Einstein equation for a flat FLRW universe.

The conservation of the matter stress-energy tensor \eqref{eq:stress_energy_conservation} for the perfect fluid \eqref{eq:perfect_fluid_stress_tensor} gives the usual continuity equation,
\begin{align}
    \bar\rho'+3\mathcal H(\bar\rho+\bar p)=0~.
    \label{eq:background_continuity}
\end{align}
Together with a specified matter model, Eqs.~\eqref{eq:modified_raychaudhuri} and \eqref{eq:background_continuity} determine the background evolution. These are the equations we use below for the slow-roll inflationary background.

The phase-space analysis of Ref.~\cite{decesare2023cosmologicalevolutionmodifiedbekenstein} shows that the modified dynamics admits self-accelerating branches at both early and late times, but the early-time branch is not a substitute for conventional inflation. It neither resolves the initial singularity through a bounce nor produces slow-roll expansion per se---unless the matter sector already has an equation of state close to $w=-1$\,. The $\alpha$-dependent term is therefore better understood as modifying an inflationary phase driven by a separate matter sector. On branches that approach the standard cosmological evolution, its contribution becomes subleading at late times.

\subsection{Linear perturbations around FLRW}
\label{sec:linear_perturbations_review}

We now turn to linear perturbations around the spatially flat FLRW background. The equations collected below were derived in Ref.~\cite{decesare2023cosmologicalevolutionmodifiedbekenstein} and provide the starting point for the inflationary calculation in Section~\ref{sec:modified_ms_equation}.

\subsubsection{Tensor perturbations}
\label{sec:tensor_perturbations_review}

We introduce tensor perturbations through
\begin{align}
    \dd s^2=a^2(\eta)\left[- \dd\eta^2+\left(\delta_{ij}+h_{ij}\right)\dd x^i\dd x^j\right]~,\label{eq:tensor_perturbed_metric}
\end{align}
where $h_{ij}$ is assumed transverse and traceless
\begin{align}
    \partial_i h^i{}_j=0~,\qquad h^i{}_i=0~.
\end{align}
Denoting the anisotropic stress component of the perturbed matter stress-energy tensor by $\pi_{ij}$, the linearised field equations read
\begin{align}
    \left[1+\frac{\alpha\kappa}{a^2}\left(\mathcal H'-\mathcal H^2\right)\right]\left(h_{ij}''+2\mathcal H h_{ij}'-\Delta h_{ij}\right)=2\kappa a^2 \pi_{ij}~.\label{eq:tensor_perturbation_equation}
\end{align}
The quantum-gravity correction therefore appears as a time-dependent prefactor multiplying the usual tensor wave operator. In the absence of anisotropic stress, and provided this prefactor does not vanish, the equation reduces to
\begin{align}
    h_{ij}''+ 2\mathcal H h_{ij}' - \Delta h_{ij}=0~.
    \label{eq:tensor_equation_no_anisotropic_stress}
\end{align}
Free tensor modes thus satisfy the same propagation equation as in general relativity. Their evolution can nevertheless differ because the scale factor and Hubble rate obey the modified background equation \eqref{eq:modified_raychaudhuri}.

\subsubsection{Scalar perturbations}
\label{sec:scalar_perturbations_review}

In longitudinal gauge, scalar perturbations are described by
\begin{align}
    \dd s^2=a^2(\eta)\left[-(1+2\Phi)\dd\eta^2+(1-2\Psi)\delta_{ij}\dd x^i\dd x^j\right]~.\label{eq:scalar_perturbed_metric}
\end{align}
In this gauge, $\Phi$ and $\Psi$ coincide with the gauge-invariant Bardeen potentials. The matter perturbations are parametrised as
\begin{align}
    \delta T^0{}_0 & =-\delta\rho~ , & \delta T^0{}_i & =(\bar\rho+\bar p)v_{,i}~ , & \delta T^i{}_j & =\delta p\,\delta^i{}_j~,    \label{eq:matter_scalar_perturbations}
\end{align}
where $v$ is the scalar velocity potential. Linearising Eq.~\eqref{eq:entropy_corrected_field_equations} gives the following independent scalar equations:\footnote{Equation~($4.4$c) of
Ref.~\cite{decesare2023cosmologicalevolutionmodifiedbekenstein} contains a typographical error: the term $2(\mathcal H'-\mathcal H^2)\Phi'$ should read $2(\mathcal H'-\mathcal H^2)\Phi$\,, as written in Eq.~\eqref{eq:scalar_dynamical_equation}.}
\begin{subequations}
\begin{align}
    \Psi-\Phi&=0~,\label{eq:scalar_no_slip}\\
    \left[1+\frac{\alpha\kappa}{a^2}\left(\mathcal H^2-\mathcal H'\right)\right]\left(\Phi'+\mathcal H\Phi \right)&=-\frac{\kappa}{2}a^2(\bar\rho+\bar p)v~,\label{eq:scalar_momentum_constraint}\\
    \left[1+\frac{2\alpha\kappa}{a^2}\left(\mathcal H^2-\mathcal H'\right)\right]\left[\Phi''+2\left(\mathcal H'-\mathcal H^2\right)\Phi+\Delta\Phi\right]&=\frac{\kappa}{2}a^2(\delta\rho+\delta p)~.\label{eq:scalar_dynamical_equation}
\end{align}
\end{subequations}
The first equation gives $\Phi=\Psi$\,, as in general relativity in the absence of matter anisotropic stress. In the remaining equations, the quantum-gravity correction enters through time-dependent prefactors multiplying the usual momentum and dynamical constraints. When the matter sector is specialised to a single scalar field, these factors combine into the effective Mukhanov--Sasaki background function derived in Section~\ref{sec:modified_ms_equation}.

\section{Scalar perturbations during inflation}
\label{sec:modified_ms_equation}

We now specialise the perturbation equations reviewed above to a single-field matter sector. The scalar field provides the slowly evolving quasi-de Sitter background, while the metric perturbations are governed by the quantum-gravity corrected field equations. The goal is to obtain a closed equation for the comoving curvature perturbation.

\subsection{Single-field background and perturbations}
\label{sec:single_field_setup}

We take the matter sector to be a single minimally coupled scalar field $\varphi$ with an arbitrary potential $V(\varphi)$. The field obeys the Klein--Gordon equation \cite{1992PhR...215..203M,riotto2017inflationtheorycosmologicalperturbations,baumann2012tasilecturesinflation}
\begin{align} 
\Box\varphi - V_{,\varphi} = 0~. \label{eq:scalar_field_KG} 
\end{align}
We decompose it into a homogeneous background and an inhomogeneous perturbation,
\begin{align}
    \varphi(\eta,\bm x) = \bar\varphi(\eta)+\delta\varphi(\eta,\bm x)~.
    \label{eq:scalar_field_split}
\end{align}
The background energy density and pressure are 
\begin{align} 
\bar\rho &= \frac{\bar\varphi'^2}{2a^2} + V(\bar\varphi)~, & \bar p &= \frac{\bar\varphi'^2}{2a^2} - V(\bar\varphi)~, & \bar\rho+\bar p &= \frac{\bar\varphi'^2}{a^2}~. \label{eq:scalar_background_rho_p} \end{align}
The background Klein--Gordon equation becomes
\begin{align}
    \bar\varphi''+2\mathcal H\bar\varphi'+a^2 V_{,\varphi}=0~.
    \label{eq:background_KG}
\end{align}
Substituting $\bar\rho+\bar p=\bar\varphi'^2/a^2$ into the modified Raychaudhuri equation \eqref{eq:modified_raychaudhuri} gives
\begin{align}
    \left( \mathcal H^2-\mathcal H' \right)\left[1+\frac{\alpha\kappa}{a^2}\left(\mathcal H^2-\mathcal H'\right)\right]=\frac{\kappa}{2}\bar\varphi'^2~.
    \label{eq:inflation_background_identity}
\end{align}
This relation will be useful when simplifying the perturbation equations. At linear order, the Klein--Gordon equation in longitudinal gauge reads \cite{baumann2012tasilecturesinflation}
\begin{align}
    \delta\varphi''+2\mathcal H\,\delta\varphi'-\nabla^2\delta\varphi-4\Phi'\bar\varphi'+a^2V_{,\varphi\varphi}\delta\varphi+2a^2V_{,\varphi}\Phi=0~.
    \label{eq:perturbed_KG}
\end{align}
Together with the scalar metric equations \eqref{eq:scalar_no_slip}--\eqref{eq:scalar_dynamical_equation}, Eqs.~\eqref{eq:background_KG} and \eqref{eq:perturbed_KG} determine the dynamics of scalar perturbations during inflation.

\subsection{Comoving curvature perturbation}
\label{sec:curvature_perturbation}

The gauge-invariant comoving curvature perturbation is defined as
\begin{align}
    \mathcal R =\Phi+\frac{\mathcal H}{\bar\varphi'}\delta\varphi~.
    \label{eq:curvature_perturbation_definition}
\end{align}
This is the dynamical variable whose late-time two-point function determines the scalar primordial spectrum \cite{1992PhR...215..203M,Malik_2009,riotto2017inflationtheorycosmologicalperturbations,baumann2012tasilecturesinflation}. We now derive a closed equation for each Fourier mode $\mathcal R_{\bm k}$, with $\Delta\to-k^2$. For a scalar-field source, the momentum constraint \eqref{eq:scalar_momentum_constraint} gives
\begin{align}
    \delta\varphi_{\bm k} = \frac{2} {\kappa\bar\varphi'}\left[1+\frac{\alpha\kappa}{a^2}\left(\mathcal H^2-\mathcal H' \right)\right]\left(\Phi_{\bm k}'+\mathcal H\Phi_{\bm k}\right)~.
    \label{eq:deltavarphi_from_momentum_constraint}
\end{align}
Substituting this expression into Eq.~\eqref{eq:curvature_perturbation_definition} and using the background identity \eqref{eq:inflation_background_identity}, one obtains
\begin{align}
    \mathcal R_{\bm k} =\Phi_{\bm k}+\frac{\mathcal H}{\mathcal H^2-\mathcal H'}\left( \Phi_{\bm k}'+\mathcal H\Phi_{\bm k}\right)~,
    \label{eq:R_in_terms_of_Phi}
\end{align}
or, equivalently,
\begin{align}
    \Phi_{\bm k}'=  \frac{\mathcal H^2-\mathcal H'}{\mathcal H}\mathcal R_{\bm k}-\left(2\mathcal H-\frac{\mathcal H'}{\mathcal H} \right)\Phi_{\bm k}~.
    \label{eq:Phiprime_RPhi}
\end{align}
Differentiating Eq.~\eqref{eq:curvature_perturbation_definition} gives
\begin{align}
    \delta\varphi_{\bm k}' = \left( \mathcal R_{\bm k}' -  \Phi_{\bm k}'
    \right) \frac{\bar\varphi'}{\mathcal H}+\left(\frac{\bar\varphi''}{\bar\varphi'}-\frac{\mathcal H'}{\mathcal H}\right)\delta\varphi_{\bm k}~.
    \label{eq:deltavarphi_prime_from_R}
\end{align}
We now combine (in Fourier space) Eq.~\eqref{eq:scalar_dynamical_equation} with the perturbed Klein--Gordon equation \eqref{eq:perturbed_KG}. The background equation \eqref{eq:background_KG} is used to remove $V_{,\varphi}$\,, while Eqs.~\eqref{eq:deltavarphi_from_momentum_constraint} and \eqref{eq:deltavarphi_prime_from_R} are used to eliminate $\delta\varphi_{\bm k}$ and $\delta\varphi_{\bm k}'$\,. After some straightforward algebraic simplifications, the resulting equation reads
\begin{align}
    \mathcal R_{\bm k}'' +2\left(\mathcal H-\frac{\mathcal H'}{\mathcal H}+\frac{\bar\varphi''}{\bar\varphi'}\right)\mathcal R_{\bm k}'+k^2\mathcal R_{\bm k}+k^2\frac{2\alpha\kappa}{a^2}\left(\mathcal H^2-\mathcal H'\right) \Phi_{\bm k}= 0~.
    \label{eq:R_equation_partially_closed}
\end{align}
The first three terms have the standard single-field structure. The last term is the explicit quantum-gravity correction and still contains the metric potential $\Phi_{\bm k}$\,. Thus Eq.~\eqref{eq:R_equation_partially_closed} is not yet a closed dynamical equation in terms of $\mathcal R_{\bm k}$ alone.

Using Eqs.~\eqref{eq:R_in_terms_of_Phi} and \eqref{eq:Phiprime_RPhi}, together with the scalar perturbation equations, we can eliminate the remaining metric potential $\Phi_{\bm k}$\,. On the branch that is continuous as $\alpha\to0$, the result is
\begin{align}
    \Phi_{\bm k} =-\frac{a^2\kappa\bar\varphi'^2}{ \mathcal H k^2 \sqrt{a^2+2\alpha\kappa^2\bar\varphi'^2}\left(\sqrt{a^2+2\alpha\kappa^2\bar\varphi'^2}
             + a \right)}\mathcal R_{\bm k}'~.
    \label{eq:Phi_in_terms_of_Rprime}
\end{align}
Substituting Eq.~\eqref{eq:Phi_in_terms_of_Rprime} into Eq.~\eqref{eq:R_equation_partially_closed} gives the desired dynamical equation for the comoving curvature perturbation
\begin{align}
    \mathcal R_{\bm k}'' +\left[2 \left( \mathcal H - \frac{\mathcal H'}{\mathcal H}+\frac{\bar\varphi''}{\bar\varphi'}\right)-\Gamma_{\rm ent}(\eta)\right]\mathcal R_{\bm k}'+k^2\mathcal R_{\bm k} = 0~,
    \label{eq:R_equation_closed}
\end{align}
where, for compactness, we have defined
\begin{align}
    \Gamma_{\rm ent}(\eta) \coloneqq \frac{\left(   \mathcal H^2-\mathcal H' \right)\left(\sqrt{a^2+2\alpha\kappa^2\bar\varphi'^2}-a \right) }{\mathcal H \sqrt{a^2+2\alpha\kappa^2\bar\varphi'^2}}~.
    \label{eq:C_eta_definition}
\end{align}
The symbol $\Gamma$ reflects its role as an additional contribution to the friction coefficient, while the subscript ``ent'' records its origin in the logarithmic correction to the horizon entropy. This correction term vanishes for $\alpha\to0$\,, and Eq.~\eqref{eq:R_equation_closed} reduces smoothly to the standard equation for the comoving curvature perturbation.

\subsection{Modified Mukhanov--Sasaki equation}
\label{sec:modified_mukhanov_sasaki_equation}

Equation~\eqref{eq:R_equation_closed} shows that the quantum-gravity correction modifies the friction term in the equation for $\mathcal R_{\bm k}$\,, while the gradient term remains $k^2\mathcal R_{\bm k}$\,. To write this equation in Mukhanov--Sasaki form, we introduce the standard background function
\begin{align}
    z_{\rm GR}   \coloneqq    \frac{a\bar\varphi'}{\mathcal H},    \qquad    \frac{z_{\rm GR}'}{z_{\rm GR}}    =    \mathcal H - \frac{\mathcal H'}{\mathcal H}+\frac{\bar\varphi''}{\bar\varphi'}~.
    \label{eq:zGR_definition}
\end{align}
The curvature perturbation equation can then be written as
\begin{align}
    \mathcal R_{\bm k}''+  \left(2\frac{z_{\rm GR}'}{z_{\rm GR}}  - \Gamma_{\rm ent}\right) \mathcal R_{\bm k}' + k^2\mathcal R_{\bm k} = 0~.
    \label{eq:R_equation_zGR_form}
\end{align}
We absorb the modified friction term into an effective Mukhanov--Sasaki background function $z_{\rm eff}$, defined by
\begin{align}
    \frac{z_{\rm eff}'}{z_{\rm eff}}  &= \frac{z_{\rm GR}'}{z_{\rm GR}} - \frac12\Gamma_{\rm ent}(\eta) = \mathcal H-\frac{\mathcal H'}{\mathcal H}  + \frac{\bar\varphi''}{\bar\varphi'}  - \frac12\Gamma_{\rm ent}(\eta)~. \label{eq:zeff_prime_over_zeff}
\end{align}
Defining $v_{\bm k}  =  z_{\rm eff}\mathcal R_{\bm k}$\,, Eq.~\eqref{eq:R_equation_closed} becomes
\begin{align}
    v_{\bm k}''  +  \left( k^2 - \frac{z_{\rm eff}''}{z_{\rm eff}}\right) v_{\bm k} = 0~.  \label{eq:modified_MS_equation}
\end{align}
Equation~\eqref{eq:zeff_prime_over_zeff} determines $z_{\rm eff}$ only up to an overall constant factor. This factor cancels out of $z_{\rm eff}''/z_{\rm eff}$ and therefore does not affect the mode equation. It does, however, enter the relation $\mathcal R_{\bm k}=v_{\bm k}/z_{\rm eff}$\,. Once the usual Bunch--Davies condition is imposed on $v_{\bm k}$ at short wavelengths, a constant rescaling $z_{\rm eff}\to C z_{\rm eff}$ gives $\mathcal R_{\bm k}\to C^{-1}\mathcal R_{\bm k}$\,. It therefore changes the overall amplitude of the scalar spectrum, while leaving its $k$ dependence unchanged.

In a theory derived from a quadratic action, the canonical normalisation would fix this freedom. Since no such action is presently known for the modified gravitational dynamics, we choose a reference time $\eta_\diamond$\,, associated below with the pivot scale, and impose
\begin{align}
    z_{\rm eff}(\eta_\diamond) =   z_{\rm GR}(\eta_\diamond)~. \label{eq:zeff_pivot_normalization}
\end{align}
This gives
\begin{align}
    z_{\rm eff}(\eta) =z_{\rm GR}(\eta)\exp\left[-\frac12\int_{\eta_\diamond}^{\eta} \Gamma_{\rm ent}(\tilde\eta)\,\dd\tilde\eta\right]~.   \label{eq:zeff_definition}
\end{align}
This prescription removes the arbitrary constant in the definition of $z_{\rm eff}$\,. It does not fix the numerical value of the scalar amplitude, which still depends on the inflationary background and is computed in Section~\ref{sec:scalar_power_spectrum}.

In the $\alpha\to0$ limit, we have $\Gamma_{\rm ent}\to0$ and therefore $z_{\rm eff}\to z_{\rm GR}$\,, so the standard Mukhanov--Sasaki equation is recovered \cite{1992PhR...215..203M}. The quantum-gravity correction therefore modifies $z_{\rm eff}''/z_{\rm eff}$\,, while leaving the sound speed (i.e., the prefactor of $k^2$) equal to one, as in GR.

\section{Primordial power spectra at
\texorpdfstring{$\mathrm{N}^3\mathrm{LO}$}{N3LO}}
\label{sec:N3LO_power_spectra}

With the effective Mukhanov--Sasaki equation in hand, we now compute the primordial scalar and tensor spectra. Our aim is to follow the $\alpha$-dependent modification through the slow-roll expansion and identify the order at which it first affects the amplitude, tilt, runnings, tensor-to-scalar ratio, and single-field consistency relation.

We describe the slow evolution of the inflationary background using the Hubble-flow hierarchy,
\begin{align}
    \epsilon_1 \coloneqq -\frac{\dd\ln H}{\dd N}~, \qquad \epsilon_{i+1}\coloneqq \frac{\dd\ln|\epsilon_i|}{\dd N}~,    \label{eq:HFF_def_sec4}
\end{align}
where $N=\ln a$ is the number of e-folds. It is convenient to define
\begin{align}
    \delta(\eta)   \coloneqq  \alpha\kappa H^2(\eta)~.
    \label{eq:delta_alpha_def}
\end{align}
The explicit perturbation-sector factors depend on the slow-roll-suppressed combination
\begin{equation*}
    \delta\epsilon_1 =\alpha\kappa H^2\epsilon_1 =-\alpha\kappa\dot H~.
\end{equation*}
We assume $|\delta\epsilon_1|\ll1$ and retain only terms linear in this combination. The Hubble rate and Hubble-flow functions, however, are evaluated on a background satisfying the modified cosmological equations \eqref{eq:modified_raychaudhuri}, without expanding that background solution in powers of $\alpha$\,. This is therefore a partially resummed organisation of the calculation: it isolates the leading explicit correction in the perturbation sector while retaining the modified background. It is not a global expansion of every observable to first order in $\alpha$\,. For a strictly order-by-order calculation in $\alpha$\,, one must additionally expand $H$ and the Hubble-flow functions about their general-relativistic values for a specified inflationary potential.

The explicit correction to the scalar mode equation is itself slow-roll suppressed, which is why it first affects the scale-dependent observables beyond leading order. Using
\begin{align}
    \mathcal H^2-\mathcal H'  =  -a^2\dot H  =  a^2H^2\epsilon_1~,  \label{eq:Hcal_identity_epsilon}
\end{align}
the correction to the friction term can be written exactly as
\begin{align}
    \Gamma_{\rm ent} =\frac{2\delta\,\epsilon_1^2\mathcal H} {1+2\delta\epsilon_1} =2\delta\,\epsilon_1^2\mathcal H +\mathcal O\!\left[\epsilon_1\mathcal H(\delta\epsilon_1)^2\right]~.
    \label{eq:gamma_slow_roll_expansion}
\end{align}
It therefore vanishes in exact de Sitter spacetime and contributes only when the background evolves away from it. To capture its first effects on the scalar tilt and its runnings, we carry the slow-roll expansion through next-to-next-to-next-to-leading order, $\mathrm{N}^3\mathrm{LO}$\,, following the pivot-expanded Green's function calculation of Ref.~\cite{Auclair_2022}.

We treat Eq.~\eqref{eq:modified_MS_equation} as an effective mode equation. In standard single-field inflation, the quadratic action identifies the canonically normalised Mukhanov--Sasaki variable and fixes the normalisation of its modes. No corresponding quadratic action is presently known for the modified gravitational dynamics considered here. Since the short-wavelength limit of Eq.~\eqref{eq:modified_MS_equation} remains $v_k''+k^2v_k \simeq 0$ as in general relativity, we adopt the usual Bunch--Davies plane-wave normalisation for $v_k$ in the subhorizon regime. The remaining multiplicative freedom in $z_{\rm eff}$ is fixed by the pivot prescription \eqref{eq:zeff_pivot_normalization}. Both choices are normalisation prescriptions within the effective mode-equation description.

The Green's function method separates the mode equation into the exactly solvable problem in  de~Sitter and a time-dependent source that measures the departure of $z_{\rm eff}''/z_{\rm eff}$ from its de~Sitter form \cite{Stewart_2001,Auclair_2022}. Because the coefficient of $k^2$ is unchanged, the homogeneous solution and the Green's function kernels are the same as in the standard calculation. The modified dynamics is carried instead by the source and by the corrected background quantities that enter its slow-roll expansion.

\subsection{Scalar spectrum from the Green's function expansion}
\label{sec:scalar_spectrum_GSR}

For each Fourier mode, we introduce the dimensionless variables
\begin{align}
    x \coloneqq -k\eta>0\,, \qquad y(x) \coloneqq \sqrt{2k}\,v_k(\eta)~. \label{eq:x_y_definitions}
\end{align}
The modified Mukhanov--Sasaki equation then becomes
\begin{align}
    \frac{\dd^2 y}{\dd x^2} + \left[ 1 - \frac{1}{k^2} \frac{z_{\rm eff}''}{z_{\rm eff}} \right]y = 0~. \label{eq:y_equation_zeff}
\end{align}
To separate the exact de Sitter evolution from the slow variation of the background, we define
\begin{align}
    g(\ln x) \coloneqq \eta^2 \frac{z_{\rm eff}''}{z_{\rm eff}} - 2~. \label{eq:gdef_main}
\end{align}
The mode equation can then be written as
\begin{align}
    \frac{\dd^2 y}{\dd x^2} + \left( 1-\frac{2}{x^2} \right)y = \frac{g(\ln x)}{x^2}y~. \label{eq:GSR_main}
\end{align}
Setting $g=0$ gives the exactly solvable de Sitter equation, whose Bunch--Davies solution is
\begin{align}
    y_0(x) = \left( 1+\frac{i}{x}\right)e^{ix}~. \label{eq:y0}
\end{align}
All departures from exact de Sitter evolution, including the explicit quantum-gravity correction, are contained in the source function $g$\,.

Using $y_0$ and $y_0^*$, Eq.~\eqref{eq:GSR_main} can be written as the retarded integral equation
\begin{align}
    y(x) = y_0(x) + \int_x^\infty \frac{\dd u}{u^2} G(x,u)\, g(\ln u)\, y(u)~, \label{eq:integral_eq}
\end{align}
where
\begin{align}
    G(x,u) = \frac{i}{2} \left[y_0^*(u)y_0(x)  - y_0(u)y_0^*(x) \right]\Theta(u-x)~. \label{eq:green}
\end{align}
For a slowly evolving background, the source can be expanded around the mode-dependent time defined by $x_\flat=1$\,,
\begin{align}
    g(\ln x) = g_{1\flat} + g_{2\flat}\ln x + g_{3\flat}\ln^2 x + \mathcal O(\epsilon^4)~, \label{eq:g_local_main}
\end{align}~
where the subscript $\flat$ denotes evaluation at that time.

We solve Eq.~\eqref{eq:integral_eq} iteratively,
\begin{align}
    y(x) = y_0(x) + y_1(x) + y_2(x) + y_3(x) + \mathcal O(\epsilon^4)~, \label{eq:y_iterative_expansion}
\end{align}
with
\begin{subequations}
\label{eq:y123_main}
\begin{align}
    y_1(x) &= g_{1\flat} \int_x^\infty \frac{\dd u}{u^2} G(x,u)y_0(u)~, \label{eq:y1_main} \\
    y_2(x) &= g_{2\flat} \int_x^\infty \frac{\dd u}{u^2} G(x,u)\ln u\,y_0(u) + g_{1\flat} \int_x^\infty \frac{\dd u}{u^2} G(x,u)y_1(u)~, \label{eq:y2_main} \\
    y_3(x) &= g_{3\flat} \int_x^\infty \frac{\dd u}{u^2} G(x,u)\ln^2u\,y_0(u) + g_{2\flat} \int_x^\infty \frac{\dd u}{u^2} G(x,u)\ln u\,y_1(u) \nonumber \\
    &\quad + g_{1\flat} \int_x^\infty \frac{\dd u}{u^2} G(x,u)y_2(u)~. \label{eq:y3_main}
\end{align}
\end{subequations}
We evaluate these integrals following Ref.~\cite{Auclair_2022}. The Green's function kernels are unchanged from the standard calculation, while the correction to the mode evolution is encoded in the source coefficients $g_{n\flat}$\,. The final scalar spectrum further contains the factor $z_{\rm eff}^{-2}$\,, whose contribution also needs to be included when computing the late-time spectrum.

\subsection{Quantum-gravity corrections to the source coefficients}
\label{sec:entropy_corrected_source_coefficients}

We expand $z_{\rm eff}''/z_{\rm eff}$ to third order in the Hubble-flow functions and to first order in the explicit quantum-gravity correction. The derivation is given in Appendix~\ref{app:gcoefficients}. Evaluating the result at the mode-dependent time $x_\flat=1$\,, we define
\begin{align}
    \delta_\flat  \coloneqq \alpha\kappa H_\flat^2~, \label{eq:delta_flat_def}
\end{align}
where $H_\flat$ is the Hubble rate at that time. The source coefficients are
\begingroup
\small
\allowdisplaybreaks
\setlength{\jot}{3pt}
\begin{subequations}
\label{eq:g_coefficients_main}
\begin{align}
g_{1\flat} &=
\delta_\flat
\left[
    -6\epsilon_{1\flat}^3 -6\epsilon_{1\flat}^2
    -\frac{9}{2}\epsilon_{2\flat}\epsilon_{1\flat}^2
    +\epsilon_{1\flat}\epsilon_{2\flat}^2
    +\frac{3}{2}\epsilon_{1\flat}\epsilon_{2\flat}
    +\frac{1}{2}\epsilon_{1\flat}\epsilon_{2\flat}\epsilon_{3\flat}
\right] +5\epsilon_{1\flat}^3 +\frac{35}{2}\epsilon_{2\flat}\epsilon_{1\flat}^2
+4\epsilon_{1\flat}^2 \nonumber\\
&\quad +\frac{15}{2}\epsilon_{2\flat}^2\epsilon_{1\flat}
+\frac{13}{2}\epsilon_{2\flat}\epsilon_{1\flat}
+5\epsilon_{2\flat}\epsilon_{3\flat}\epsilon_{1\flat} +3\epsilon_{1\flat} +\frac{1}{4}\epsilon_{2\flat}^2
+\frac{3}{2}\epsilon_{2\flat} +\frac{1}{2}\epsilon_{2\flat}\epsilon_{3\flat}~,
\label{eq:g1_main}
\\[0.5em]
g_{2\flat} &=
\delta_\flat\left[ -12\epsilon_{1\flat}^3    +15\epsilon_{2\flat}\epsilon_{1\flat}^2 -\frac{3}{2}\epsilon_{1\flat}\epsilon_{2\flat}^2  -\frac{3}{2}\epsilon_{1\flat}\epsilon_{2\flat}\epsilon_{3\flat}
\right] -11\epsilon_{2\flat}\epsilon_{1\flat}^2
-\frac{13}{2}\epsilon_{2\flat}^2\epsilon_{1\flat}-3\epsilon_{2\flat}\epsilon_{1\flat}\nonumber\\
&\quad-8\epsilon_{2\flat}\epsilon_{3\flat}\epsilon_{1\flat} -\frac{1}{2}\epsilon_{2\flat}\epsilon_{3\flat}^2 -\frac{1}{2}\epsilon_{2\flat}^2\epsilon_{3\flat}-\frac{3}{2}\epsilon_{2\flat}\epsilon_{3\flat}-\frac{1}{2}\epsilon_{2\flat}\epsilon_{3\flat}\epsilon_{4\flat}~,
\label{eq:g2_main}
\\[0.5em]
g_{3\flat}
&=\frac{3}{2}\epsilon_{1\flat}\epsilon_{2\flat}^2
+\frac{3}{4}\epsilon_{3\flat}^2\epsilon_{2\flat}
+\frac{3}{2}\epsilon_{1\flat}\epsilon_{3\flat}\epsilon_{2\flat}
+\frac{3}{4}\epsilon_{3\flat}\epsilon_{4\flat}\epsilon_{2\flat}~.
\label{eq:g3_main}
\end{align}
\end{subequations}
\endgroup
In the $\alpha\to0$ limit, $\delta_\flat$ vanishes, and thus one recovers the standard single-field coefficients, see Eqs.~(B.10)--(B.12) of Ref.~\cite{Auclair_2022}. At the order considered here, the explicit quantum-gravity correction contributes to $g_{1\flat}$ and $g_{2\flat}$, while $g_{3\flat}$ contains no term proportional to $\delta_\flat$\,.

\subsection{Scalar power spectrum}
\label{sec:scalar_power_spectrum}

The scalar power spectrum is determined by the late-time value of
$\mathcal R_k=v_k/z_{\rm eff}$\,:
\begin{align}
    \mathcal P_{\mathcal R}(k) = \frac{k^3}{2\pi^2} \left| \frac{v_k}{z_{\rm eff}} \right|_{x\to0}^{2}=\frac{1}{4\pi^2}\lim_{x\to0^+}\left|x\,y(x)\right|^2\left|\frac{k}{x z_{\rm eff}}\right|^2~.\label{eq:PR_definition_GSR}
\end{align}
The first factor in Eq.~\eqref{eq:PR_definition_GSR} is obtained from the Green's function solution. After expanding the source as in Eq.~\eqref{eq:g_local_main}, the required integrals depend only on the de Sitter mode function $y_0$, the Green's function $G(x,u)$, and powers of $\ln u$. For example, the first iteration contains integrals of the form
\begin{align}
    \lim_{x\to0} x \int_x^\infty \frac{\dd u}{u^2} G(x,u) (\ln u)^n y_0(u)~,
    \label{eq:universal_GSR_integral_example}
\end{align}
with nested integrals appearing at higher orders. Their values are universal numerical constants \cite{Auclair_2022}. The inflationary background and the quantum-gravity correction enter through the source coefficients $g_{n\flat}$\,. The second factor in Eq.~\eqref{eq:PR_definition_GSR} contains the corresponding correction from $z_{\rm eff}^{-2}$\,. Both contributions need to be included in the final expression.

The source coefficients are evaluated at the mode-dependent time $x_\flat=1$\,. Observational predictions, however, are conventionally expressed around a fixed pivot scale, which we take to be $k_\diamond=0.05\,{\rm Mpc}^{-1}$\,. We therefore re-expand $H_\flat$, $\delta_\flat$\,, and the Hubble-flow functions $\epsilon_{i\flat}$ around the time at which the pivot mode satisfies $x_\diamond=1$\,. This converts the result into a polynomial in $L \coloneqq \ln\left(   \frac{k}{k_\diamond}\right)$\,. The required relations between the mode-dependent and pivot quantities are collected in Appendix~\ref{app:GSR_details}.

Using the normalisation prescription
\eqref{eq:zeff_pivot_normalization}, the scalar spectrum around the
pivot takes the form
\begin{align}
    \mathcal P_{\mathcal R}(k) = \frac{\kappa H_\diamond^2} {8\pi^2\epsilon_{1\diamond}}\left[ P_0+ P_1 L + P_2 L^2 + P_3 L^3\right] + \mathcal O(\epsilon^4)~.
    \label{eq:PR_main}
\end{align}
We define
\begin{align}
    \mathcal C \coloneqq  \gamma_E+\ln2-2~,  \qquad \delta_\diamond  \coloneqq    \alpha\kappa H_\diamond^2~,
    \label{eq:C_delta_diamond_def}
\end{align}
where $\gamma_E$ is the Euler--Mascheroni constant. We also denote the Riemann zeta function evaluated at $3$ by $\zeta(3)$\,.\\

\noindent The $\mathrm{N}^3\mathrm{LO}$ coefficients are
\begingroup
\small
\allowdisplaybreaks
\setlength{\jot}{3pt}
\begin{subequations}
\label{eq:P_coefficients_main}
\begin{align}
P_0 &=1-2\left(  \mathcal C+1+\frac{\delta_\diamond}{2}\right)\epsilon_{1\diamond}-\mathcal C\epsilon_{2\diamond}+\left[\frac{\pi^2}{2} +2\mathcal C^2 + 2\mathcal C -3+\delta_\diamond(6\mathcal C+2)\right]\epsilon_{1\diamond}^2 \nonumber\\
&\quad
+\left( \frac{7\pi^2}{12} +\mathcal C^2 -\mathcal C -6 \right)\epsilon_{1\diamond}\epsilon_{2\diamond} +\frac18 \left( \pi^2+4\mathcal C^2-8 \right)\epsilon_{2\diamond}^2 +\frac1{24}
\left(
    \pi^2-12\mathcal C^2
\right)\epsilon_{2\diamond}\epsilon_{3\diamond}\nonumber \\
&\quad
-\frac1{24}
\left[    4\mathcal C^3    +3(\pi^2-8)\mathcal C    +14\zeta(3)    -16\right] \left( 8\epsilon_{1\diamond}^3+\epsilon_{2\diamond}^3 \right) \nonumber \\
&\quad
+\frac1{12}
\left[    13\pi^2    -8(\pi^2-9)\mathcal C    +36\mathcal C^2  -84\zeta(3)    -12\delta_\diamond    \left(   \frac32\pi^2 -2\mathcal C^2
        -2\mathcal C -10    \right)\right]\epsilon_{1\diamond}^2\epsilon_{2\diamond}
\nonumber
\\
&\quad
-\frac1{24}\left[  8\mathcal C^3   -15\pi^2 +6(\pi^2-4)\mathcal C -12\mathcal C^2     +100\zeta(3)    +16    -4\delta_\diamond(\pi^2-6)\right]\epsilon_{1\diamond}\epsilon_{2\diamond}^2\nonumber\\
&\quad
+\frac1{24}
\left(    \pi^2\mathcal C    -4\mathcal C^3 -8\zeta(3) +16\right) \left(    \epsilon_{2\diamond}\epsilon_{3\diamond}^2  +\epsilon_{2\diamond}\epsilon_{3\diamond}\epsilon_{4\diamond}
\right)\nonumber\\
&\quad
+\frac1{24}\left[12\mathcal C^3 +(5\pi^2-48)\mathcal C \right]\epsilon_{2\diamond}^2\epsilon_{3\diamond} -\delta_\diamond\epsilon_{1\diamond}^3 \left(    \frac{13}{6}\pi^2    +14\mathcal C^2   +6\mathcal C   -19\right)\nonumber\\
&\quad
+\frac1{12}\left[  8\mathcal C^3  +\pi^2  +6(\pi^2-12)\mathcal C -12\mathcal C^2
    -8\zeta(3)    -8\right]\epsilon_{1\diamond}\epsilon_{2\diamond}\epsilon_{3\diamond}~,
\label{eq:P0_main}
\\[0.5em]
P_1&= -2\epsilon_{1\diamond} -\epsilon_{2\diamond}
+2(2\mathcal C+1+2\delta_\diamond)\epsilon_{1\diamond}^2
+(2\mathcal C-1)\epsilon_{1\diamond}\epsilon_{2\diamond}
+\mathcal C\epsilon_{2\diamond}^2
-\mathcal C\epsilon_{2\diamond}\epsilon_{3\diamond}
\nonumber\\
&\quad
-\frac18\left(\pi^2+4\mathcal C^2-8\right)
\left(8\epsilon_{1\diamond}^3+\epsilon_{2\diamond}^3\right)
-4\delta_\diamond(6\mathcal C+1)\epsilon_{1\diamond}^3
\nonumber\\
&\quad
-\frac23
\left[
\pi^2-9\mathcal C-9
-3\delta_\diamond(3\mathcal C+1)
\right]
\epsilon_{1\diamond}^2\epsilon_{2\diamond}
-\frac14
\left(\pi^2+4\mathcal C^2-4\mathcal C-4\right)
\epsilon_{1\diamond}\epsilon_{2\diamond}^2
\nonumber\\
&\quad
+\frac12
\left(\pi^2+4\mathcal C^2-4\mathcal C-12\right)
\epsilon_{1\diamond}\epsilon_{2\diamond}\epsilon_{3\diamond}
+\frac1{24}
\left(\pi^2-12\mathcal C^2\right)
\left(
\epsilon_{2\diamond}\epsilon_{3\diamond}^2
+\epsilon_{2\diamond}\epsilon_{3\diamond}\epsilon_{4\diamond}
\right)
\nonumber\\
&\quad
+\frac1{24}
\left(5\pi^2+36\mathcal C^2-48\right)
\epsilon_{2\diamond}^2\epsilon_{3\diamond}~,
\label{eq:P1_main}
\\[0.5em]
P_2 &=
2\epsilon_{1\diamond}^2
+\epsilon_{1\diamond}\epsilon_{2\diamond}
+\frac12\epsilon_{2\diamond}^2
-\frac12\epsilon_{2\diamond}\epsilon_{3\diamond}
+\left(3+2\delta_\diamond\right)
 \epsilon_{1\diamond}^2\epsilon_{2\diamond} -\left(\mathcal C-\frac12\right)
\left(
\epsilon_{1\diamond}\epsilon_{2\diamond}^2
-2\epsilon_{1\diamond}\epsilon_{2\diamond}\epsilon_{3\diamond}
\right)
\nonumber\\
&\quad
-\left(4\mathcal C+8\delta_\diamond\right)
\epsilon_{1\diamond}^3 -\frac{\mathcal C}{2}
\left(
\epsilon_{2\diamond}^3
-3\epsilon_{2\diamond}^2\epsilon_{3\diamond}
+\epsilon_{2\diamond}\epsilon_{3\diamond}^2
+\epsilon_{2\diamond}\epsilon_{3\diamond}\epsilon_{4\diamond}
\right)~, \label{eq:P2_main} \\[0.5em]
P_3 &= \frac16 \left(     -8\epsilon_{1\diamond}^3    -2\epsilon_{1\diamond}\epsilon_{2\diamond}^2   +4\epsilon_{1\diamond}\epsilon_{2\diamond}\epsilon_{3\diamond}
    -\epsilon_{2\diamond}^3    +3\epsilon_{2\diamond}^2\epsilon_{3\diamond}    -\epsilon_{2\diamond}\epsilon_{3\diamond}^2   -\epsilon_{2\diamond}\epsilon_{3\diamond}\epsilon_{4\diamond}
\right)~.
\label{eq:P3_main}
\end{align}
\end{subequations}
\endgroup
Equations~\eqref{eq:PR_main}--\eqref{eq:P_coefficients_main} show that at $\mathrm{N}^3\mathrm{LO}$\,, the correction changes only the coefficients of the usual pivot expansion and introduces no new functional dependence on $k$\,. The details of the derivation are collected in Appendix~\ref{app:entropy_normalization}.

\subsection{Scalar observables}
\label{sec:scalar_observables}

To connect Eq.~\eqref{eq:PR_main} with the usual phenomenological parametrisation, we write
\begin{align}
    \ln\mathcal P_{\mathcal R}(k) =\ln A_s +(n_s-1)L + \frac12\alpha_s L^2  +\frac16\beta_s L^3 + \mathcal O(L^4)~,
    \label{eq:phenom_scalar_spectrum}
\end{align}
where $A_s$ is the scalar amplitude at the pivot scale, $n_s$ is the scalar spectral index, $\alpha_s$ is its running, and $\beta_s$ is the running of the running. Matching Eq.~\eqref{eq:phenom_scalar_spectrum} to Eq.~\eqref{eq:PR_main} gives
\begin{subequations}
\label{eq:scalar_observable_matching}
\begin{align}
    A_s    &=    \frac{\kappa H_\diamond^2}   {8\pi^2\epsilon_{1\diamond}}   P_0~,
    \label{eq:As_matching}    \\
    n_s-1    &=    \frac{P_1}{P_0}~,    \label{eq:ns_matching}   \\
    \alpha_s    &= 2 \left(  \frac{P_2}{P_0} - \frac12 \frac{P_1^2}{P_0^2} \right)~, \label{eq:alphas_matching}    \\
    \beta_s    &=  6 \left(\frac{P_3}{P_0}  - \frac{P_1P_2}{P_0^2}  +   \frac13
        \frac{P_1^3}{P_0^3}    \right)~.    \label{eq:betas_matching}
\end{align}
\end{subequations}
Within the scalar normalisation prescription \eqref{eq:zeff_pivot_normalization}, a specified inflationary potential can be matched to the observed value of $A_s$ to fix the overall inflationary scale. The remaining quantities describe the scale dependence of the spectrum and are insensitive to a constant rescaling of $z_{\rm eff}$.

Organising the result by slow-roll order, we write
\begin{subequations}
\label{eq:scalar_observable_expansions}
\begin{align}
    n_s  &= 1 + n_s^{(1)}  + n_s^{(2)} + n_s^{(3)} + \mathcal O(\epsilon^4)~,    \\
    \alpha_s  &= \alpha_s^{(2)} +  \alpha_s^{(3)}  + \mathcal O(\epsilon^4)~,  \\
    \beta_s &=   \beta_s^{(3)}  + \mathcal O(\epsilon^4)~.
\end{align}
\end{subequations}
The scalar tilt is
\begingroup
\small
\allowdisplaybreaks
\setlength{\jot}{3pt}
\begin{subequations}
\label{eq:ns_coefficients}
\begin{align}
    n_s^{(1)}   &=   -2\epsilon_{1\diamond} - \epsilon_{2\diamond}~,    \\
    n_s^{(2)}  &= (-2+4\delta_\diamond)\epsilon_{1\diamond}^2 -  (2\mathcal C +3+\delta_\diamond)  \epsilon_{1\diamond}\epsilon_{2\diamond}  -  \mathcal C\epsilon_{2\diamond}\epsilon_{3\diamond}~,  \\
    n_s^{(3)}   &= \left[  -2   -4\delta_\diamond(2\mathcal C-1)\right]\epsilon_{1\diamond}^3 + \left[ \pi^2 -6\mathcal C -15 +\delta_\diamond(10\mathcal C-1)\right]\epsilon_{1\diamond}^2\epsilon_{2\diamond} \nonumber \\
    &\quad  + \frac1{12} \left[ 7\pi^2 -12\mathcal C^2-36\mathcal C   -84    -12\mathcal C \delta_\diamond\right]\epsilon_{1\diamond}\epsilon_{2\diamond}^2  + \left[ \frac{7\pi^2}{12} -\mathcal C^2 -4\mathcal C -6-\mathcal C \delta_\diamond \right] \epsilon_{1\diamond}\epsilon_{2\diamond}\epsilon_{3\diamond} \nonumber   \\
    &\quad  + \frac1{24}\left( \pi^2  -12\mathcal C^2 \right) \left(      \epsilon_{2\diamond}\epsilon_{3\diamond}^2  +\epsilon_{2\diamond}\epsilon_{3\diamond}\epsilon_{4\diamond}\right)+\left(  \frac{\pi^2}{4}  -2\right) \epsilon_{2\diamond}^2\epsilon_{3\diamond}~.
\end{align}
\end{subequations}
\endgroup
The running and running of the running are
\begingroup
\small
\allowdisplaybreaks
\setlength{\jot}{3pt}
\begin{subequations}
\label{eq:running_coefficients}
\begin{align}
    \alpha_s^{(2)}    &= -2\epsilon_{1\diamond}\epsilon_{2\diamond}
    -
    \epsilon_{2\diamond}\epsilon_{3\diamond}~,   \\
    \alpha_s^{(3)}    &= -8\delta_\diamond\epsilon_{1\diamond}^3 + (-6+10\delta_\diamond)\epsilon_{1\diamond}^2\epsilon_{2\diamond}
       -   (2\mathcal C+3+\delta_\diamond)
    \epsilon_{1\diamond}\epsilon_{2\diamond}^2
      -   (2\mathcal C+4+\delta_\diamond)
    \epsilon_{1\diamond}\epsilon_{2\diamond}\epsilon_{3\diamond}
    \nonumber
    \\
    &\quad  -  \mathcal C
    \left(        \epsilon_{2\diamond}\epsilon_{3\diamond}^2        +
        \epsilon_{2\diamond}\epsilon_{3\diamond}\epsilon_{4\diamond}
    \right)~,    \\
    \beta_s^{(3)}
    &=    -2\epsilon_{1\diamond}\epsilon_{2\diamond}^2
    -    2\epsilon_{1\diamond}\epsilon_{2\diamond}\epsilon_{3\diamond}  -    \epsilon_{2\diamond}\epsilon_{3\diamond}^2
    -    \epsilon_{2\diamond}\epsilon_{3\diamond}\epsilon_{4\diamond}~.
\end{align}
\end{subequations}
\endgroup
The order at which these corrections appear can be understood simply as follows. Since $\Gamma_{\rm ent}\propto\mathcal H^2-\mathcal H' =a^2H^2\epsilon_1$\,, the explicit quantum-gravity correction vanishes in the exact de~Sitter limit and first enters the scalar amplitude at order $\delta_\diamond\epsilon$\,. Each derivative with respect to $\ln k$ raises the slow-roll order by one; see Eq.~\eqref{eq:app_D_flow_relations}. The correction therefore first appears in the scalar tilt at order $\delta_\diamond\epsilon^2$ and in the running at order $\delta_\diamond\epsilon^3$\,. Its contribution to the running of the running begins only at order $\delta_\diamond\epsilon^4$ and hence lies beyond the $\mathrm{N}^3\mathrm{LO}$ accuracy considered here. At this order, the quantum-gravity dependence of $\beta_s$ enters only through the corrected background values of the Hubble-flow functions.

\subsection{Tensor spectrum and tensor-to-scalar ratio}
\label{sec:tensor_spectrum_r}

As discussed in Section~\ref{sec:tensor_perturbations_review}, in the absence of anisotropic stress, free tensor modes obey the same classical propagation equation as in general relativity. Equality of the classical equation does not by itself fix the quantum Wronskian or the absolute tensor amplitude. Because no quadratic action is presently known for the modified dynamics, we make the independent prescription that the tensor modes have the standard general-relativistic canonical normalisation and the usual Bunch--Davies initial condition. Conditional on this prescription, we use the standard slow-roll tensor spectrum, with the Hubble rate and Hubble-flow functions evaluated on the quantum-gravity-corrected background. Using the $\mathrm{N}^3\mathrm{LO}$ result of Ref.~\cite{Auclair_2022}, the tensor spectrum around the pivot scale is
\begin{align}
    \mathcal P_T(k)  = \frac{2\kappa H_\diamond^2}{\pi^2} \left[ T_0 + T_1 L  + T_2 L^2  +     T_3 L^3\right] + \mathcal O(\epsilon^4)~,
    \label{eq:tensor_power_spectrum_N3LO}
\end{align}
where
\begingroup
\small
\allowdisplaybreaks
\setlength{\jot}{3pt}
\begin{subequations}
\label{eq:tensor_power_coefficients}
\begin{align}
T_0&=1-2(\mathcal C+1)\epsilon_{1\diamond}+\frac12\left(  \pi^2+4\mathcal C^2+4\mathcal C-6 \right)\epsilon_{1\diamond}^2  +\frac1{12} \left(     \pi^2-12\mathcal C^2-24\mathcal C-24 \right)\epsilon_{1\diamond}\epsilon_{2\diamond} \nonumber\\
&\quad-\frac13\left[    4\mathcal C^3    +3(\pi^2-8)\mathcal C     +14\zeta(3)   -16 \right]\epsilon_{1\diamond}^3 +\frac1{12}\left[    24\mathcal C^3    +13\pi^2    +2(5\pi^2-36)\mathcal C    +36\mathcal C^2    -96\right]\epsilon_{1\diamond}^2\epsilon_{2\diamond}\nonumber\\
&\quad-\frac1{12}\left[    4\mathcal C^3    -\pi^2    -(\pi^2-24)\mathcal C    +12\mathcal C^2    +8\zeta(3)    +8\right]\left(    \epsilon_{1\diamond}\epsilon_{2\diamond}^2
    +    \epsilon_{1\diamond}\epsilon_{2\diamond}\epsilon_{3\diamond}\right)~,\label{eq:T0_coeff}\\[0.4em]
T_1&=-2\epsilon_{1\diamond}+2(2\mathcal C+1)\epsilon_{1\diamond}^2-2(\mathcal C+1)\epsilon_{1\diamond}\epsilon_{2\diamond}-\left(    \pi^2+4\mathcal C^2-8\right)\epsilon_{1\diamond}^3\nonumber\\
&\quad+\frac16\left(    5\pi^2+36\mathcal C^2+36\mathcal C-36 \right)\epsilon_{1\diamond}^2\epsilon_{2\diamond}
 +\frac1{12}\left(    \pi^2-12\mathcal C^2-24\mathcal C-24\right)\left(    \epsilon_{1\diamond}\epsilon_{2\diamond}^2    +   \epsilon_{1\diamond}\epsilon_{2\diamond}\epsilon_{3\diamond}\right),\label{eq:T1_coeff}\\[0.4em]
T_2&=2\epsilon_{1\diamond}^2-\epsilon_{1\diamond}\epsilon_{2\diamond}-4\mathcal C\epsilon_{1\diamond}^3+3(2\mathcal C+1)\epsilon_{1\diamond}^2\epsilon_{2\diamond}-(\mathcal C+1)\left(    \epsilon_{1\diamond}\epsilon_{2\diamond}^2    +    \epsilon_{1\diamond}\epsilon_{2\diamond}\epsilon_{3\diamond}\right)~,
\label{eq:T2_coeff}\\[0.4em]
T_3&=\frac13\left(    -4\epsilon_{1\diamond}^3    +6\epsilon_{1\diamond}^2\epsilon_{2\diamond} -\epsilon_{1\diamond}\epsilon_{2\diamond}^2    -\epsilon_{1\diamond}\epsilon_{2\diamond}\epsilon_{3\diamond}\right)~.
\label{eq:T3_coeff}
\end{align}
\end{subequations}
\endgroup
Within this tensor-normalisation prescription, no term proportional to $\delta_\diamond$ appears explicitly in the coefficients $T_i$. The tensor spectrum can nevertheless differ from its general-relativistic value because $H_\diamond$ and $\epsilon_{i\diamond}$ are determined by the modified background equations. A constant change of the tensor normalisation would rescale $\mathcal P_T$ but would not change the tensor tilt $n_T$.

\noindent Using the scalar normalisation fixed in Eq.~\eqref{eq:zeff_pivot_normalization}, the tensor-to-scalar ratio at
the pivot is
\begin{align}
    r    &\coloneqq    \frac{\mathcal P_T(k_\diamond)}   {\mathcal P_{\mathcal R}(k_\diamond)}
    =    16\epsilon_{1\diamond}    \frac{T_0}{P_0}~.    \label{eq:r_exact_ratio}
\end{align}
Expanding the ratio $T_0/P_0$ gives
\begin{align}
    r    =    16\epsilon_{1\diamond}    \left(  1+r^{(1)}+r^{(2)}+r^{(3)}\right)+ \mathcal O(\epsilon^5)~,    \label{eq:r_expansion}
\end{align}
where
\begingroup
\small
\allowdisplaybreaks
\setlength{\jot}{3pt}
\begin{subequations}
\label{eq:r_coefficients}
\begin{align}
r^{(1)}&=\mathcal C\epsilon_{2\diamond}+\delta_\diamond\epsilon_{1\diamond}~,
\label{eq:r1_coefficient}\\[0.4em]
r^{(2)}&=-\frac18\left(\pi^2    -4\mathcal C^2  -8\right)\epsilon_{2\diamond}^2-4\mathcal C\delta_\diamond\epsilon_{1\diamond}^2-\left(  \frac{\pi^2}{2}    -\mathcal C    -4    -2\mathcal C\delta_\diamond
\right)\epsilon_{1\diamond}\epsilon_{2\diamond}\nonumber\\
&\quad
-\frac1{24}\left(    \pi^2    -12\mathcal C^2\right)\epsilon_{2\diamond}\epsilon_{3\diamond}~,
\label{eq:r2_coefficient}
\\[0.4em]
r^{(3)}
&=\frac1{24}\left[    4\mathcal C^3-3(\pi^2-8)\mathcal C  +14\zeta(3)  -16\right]\epsilon_{2\diamond}^3+\frac13\delta_\diamond\epsilon_{1\diamond}^3
\left(    5\pi^2    +12\mathcal C^2    -12\mathcal C    -48\right)
\nonumber\\
&\quad-\frac1{24}\left(    \pi^2\mathcal C    -4\mathcal C^3    -8\zeta(3)    +16\right)
\left(    \epsilon_{2\diamond}\epsilon_{3\diamond}^2    +    \epsilon_{2\diamond}\epsilon_{3\diamond}\epsilon_{4\diamond}\right)\nonumber\\
&\quad
+\left[    \frac{\mathcal C^3}{2}    -    \frac1{24}(7\pi^2-48)\mathcal C
\right]\epsilon_{2\diamond}^2\epsilon_{3\diamond}-\left[    \pi^2    -\mathcal C    -7\zeta(3)    -    \delta_\diamond    \left(        \frac{5\pi^2}{12}
        -9\mathcal C^2        +2\mathcal C    \right)\right]\epsilon_{1\diamond}^2\epsilon_{2\diamond}
\nonumber\\
&\quad
-\frac1{12}
\left[    \pi^2    -12\mathcal C^2    +6\mathcal C(\pi^2-8)
    +    \delta_\diamond \left(    \pi^2        -12\mathcal C^2
    \right)\right]\epsilon_{1\diamond}\epsilon_{2\diamond}\epsilon_{3\diamond}
\nonumber\\
&\quad
-\frac1{24}\left[    19\pi^2    -36\mathcal C^2    +24(\pi^2-9)\mathcal C
    -84\zeta(3)    -48    +    2\delta_\diamond
    \left(     5\pi^2        -24\mathcal C^2   -36 \right)
\right]\epsilon_{1\diamond}\epsilon_{2\diamond}^2~.
\label{eq:r3_coefficient}
\end{align}
\end{subequations}
\endgroup
The first explicit quantum-gravity contribution appears in $r^{(1)}$ through the term $\delta_\diamond\epsilon_{1\diamond}$. Including the overall factor of $\epsilon_{1\diamond}$ in Eq.~\eqref{eq:r_expansion}, this changes $r$ at order $\delta_\diamond\epsilon^2$. Within the adopted normalisations, its explicit origin lies in the scalar sector: the scalar modification $z_{\rm GR}\to z_{\rm eff}$ changes $\mathcal P_{\mathcal R}$ without a corresponding explicit term in the tensor propagation equation. The numerical value of $r$ remains conditional on the relative scalar and tensor normalisations.

\subsection{Single-field consistency relation}
\label{sec:consistency_relation}

At leading order in standard single-field slow-roll inflation, the tensor-to-scalar ratio and tensor tilt satisfy
\begin{align}
    r   =   -8n_T~.    \label{eq:leading_consistency_relation}
\end{align}
Higher-order slow-roll terms modify this relation even in general relativity \cite{riotto2017inflationtheorycosmologicalperturbations, baumann2012tasilecturesinflation,Auclair_2022}. The quantum-gravity correction produces an additional change through the scalar spectrum and the modified background.

The tensor tilt follows from the tensor spectrum:
\begin{align}
    n_T     &\coloneqq  \left.   \frac{\dd\ln\mathcal P_T}    {\dd\ln k}    \right|_{k=k_\diamond}    =    \frac{T_1}{T_0}~.
    \label{eq:nT_from_tensor_coefficients}
\end{align}
For the consistency relation below, the tensor tilt is required through fourth order in the Hubble-flow functions. Although the coefficients $T_i$ in Eq.~\eqref{eq:tensor_power_coefficients} are displayed only through cubic order, the required fourth-order contribution is fixed by differentiating the $\mathrm{N}^3\mathrm{LO}$ tensor amplitude, since a derivative with respect to $\ln k$ raises the slow-roll order by one. Using $\dd/\dd\ln k=-\eta\mathcal H\,\dd/\dd N$, we therefore evaluate
\begin{align}
    n_T  &= \left. -\eta\mathcal H \frac{\dd}{\dd N} \ln\!\left(H^2T_0\right) \right|_{\diamond} +\mathcal O(\epsilon^5)~, \label{eq:nT_fourth_order}
\end{align}
This is equivalent to expanding $T_1/T_0$ while retaining the fourth-order contribution to $T_1$ generated by this differentiation. We measure the departure from the leading-order consistency relation by defining
\begin{align}
    \Delta_c     \coloneqq     r+8n_T~.
    \label{eq:Delta_c_def}
\end{align}
Combining Eq.~\eqref{eq:r_expansion} with the tensor tilt evaluated to this order in Eq.~\eqref{eq:nT_fourth_order}, we obtain
\begin{align}
    \Delta_c     =     16\epsilon_{1\diamond}     \left(  \Delta_c^{(1)}  +   \Delta_c^{(2)}   +   \Delta_c^{(3)} \right)   +  \mathcal O(\epsilon^5)~,
    \label{eq:Delta_c_expansion}
\end{align}
with
\begingroup
\small
\allowdisplaybreaks
\setlength{\jot}{3pt}
\begin{subequations}
\label{eq:Delta_c_coefficients}
\begin{align}
\Delta_c^{(1)}&=(-1+\delta_\diamond)\epsilon_{1\diamond}-\epsilon_{2\diamond}~,\label{eq:Delta_c1_coefficient}\\[0.4em]
\Delta_c^{(2)}&=(-1-4\mathcal C\delta_\diamond)\epsilon_{1\diamond}^2-(2\mathcal C+3-2\mathcal C\delta_\diamond)\epsilon_{1\diamond}\epsilon_{2\diamond}
-\left(    \frac{\pi^2}{12}    +\mathcal C\right)\epsilon_{2\diamond}^2-(\mathcal C+1)\epsilon_{2\diamond}\epsilon_{3\diamond}~,\label{eq:Delta_c2_coefficient}\\[0.4em] 
\Delta_c^{(3)}&=\left[    -1 + \delta_\diamond    \left(   \frac{5\pi^2}{3}        +4\mathcal C^2        -4\mathcal C        -16    \right)\right]\epsilon_{1\diamond}^3+
\left[    \pi^2    -5\mathcal C
    -14    +    \delta_\diamond    \left(
        \frac{5\pi^2}{12}
        -9\mathcal C^2        +2\mathcal C
    \right)
\right]\epsilon_{1\diamond}^2\epsilon_{2\diamond}\nonumber\\
&\quad+\left[
    \frac{7\pi^2}{12}  -\mathcal C^2 -4\mathcal C    -8  -
    \delta_\diamond
    \left(
        \frac{\pi^2}{12}
        -\mathcal C^2
    \right)
\right]\epsilon_{1\diamond}\epsilon_{2\diamond}\epsilon_{3\diamond}
+
\left(
    \frac{\pi^2}{24}
    -\frac{\mathcal C^2}{2}
    -\mathcal C
    -1
\right)
\left(    \epsilon_{2\diamond}\epsilon_{3\diamond}^2    +    \epsilon_{2\diamond}\epsilon_{3\diamond}\epsilon_{4\diamond}
\right)\nonumber\\
&\quad+\left[    \frac{\pi^2}{8}    -\frac{3\mathcal C^2}{2}    -\frac{\mathcal C}{6}(\pi^2+6)    -\zeta(3)    -1\right]\epsilon_{2\diamond}^2\epsilon_{3\diamond}
+\left[    \frac{\pi^2}{24}
    -\frac{\mathcal C^2}{2}
    -\frac{\mathcal C\pi^2}{12}
    +\frac{\zeta(3)}{4}
    -1
\right]\epsilon_{2\diamond}^3\nonumber\\
&\quad+\left[ \frac{\pi^2}{2} -2\mathcal C^2 -7\mathcal C +\frac{7\zeta(3)}{2}
    -13   -  \delta_\diamond \left( \frac{5\pi^2}{12}-2\mathcal C^2-3\right)
\right]\epsilon_{1\diamond}\epsilon_{2\diamond}^2~.
\label{eq:Delta_c3_coefficient}
\end{align}
\end{subequations}
\endgroup
Setting $\alpha=0$ recovers the standard higher-order slow-roll departure from $r=-8n_T$\,. For nonzero $\alpha$\,, the coefficients acquire additional quantum-gravity corrections via $\delta_\diamond$\,. Conditional on the scalar and tensor normalisation prescriptions stated above, a joint measurement of $r$ and $n_T$ would therefore test the modified consistency relation \cite{Smith_2006,Boyle_2015}. The tensor tilt itself is insensitive to a constant tensor normalisation, whereas $r$ and hence $\Delta_c$ are not.

\section{Discussion and Outlook}
\label{sec:discussion_outlook}

In this work, we studied standard single-field slow-roll inflation when the metric is governed by the modified gravitational dynamics derived from a logarithmically corrected Bekenstein--Hawking entropy law \cite{Alonso_Serrano_2020,Alonso_Serrano_2023}. We obtained a closed evolution equation for the comoving curvature perturbation and computed the scalar and tensor primordial spectra through $\mathrm{N}^3\mathrm{LO}$\,. The correction deforms the standard slow-roll predictions in two ways: it changes the inflationary background and contributes explicitly to scalar perturbations, while free tensor modes are modified only through the background.

The main physical insight comes from the scalar mode equation, Eq.~\eqref{eq:modified_MS_equation}. It retains the usual Mukhanov--Sasaki form, with the general-relativistic background function $z_{\rm GR}$ replaced by $z_{\rm eff}$\,. In particular, the coefficient of $k^2$ remains unity, so scalar modes retain the standard short-wavelength propagation speed. The correction enters through the time-dependent effective mass term $-z_{\rm eff}''/z_{\rm eff}$ and affects the amplification of the modes as they approach horizon crossing. This differs from models with non-canonical kinetic terms or more general effective descriptions of inflation, in which the scalar sound speed itself can be modified \cite{Garriga_1999,Cheung_2008,Hu_2011,MotohashiHu_2017,Bianchi_2024}.

For the smoothly evolving slow-roll backgrounds considered here, the scalar power spectrum retains its usual expansion in powers of $L=\ln(k/k_\diamond)$\,. At the perturbative order considered, the quantum-gravity correction changes the coefficients of this pivot expansion rather than introducing a new functional dependence on $k$; in particular, it produces no oscillations, sharp features, or new non-analytic scale dependence. Consequently, a nonzero $\delta=\alpha\kappa H^2$ can mimic a small change in the Hubble-flow functions. Therefore, the scalar tilt $n_s$ alone cannot cleanly distinguish the correction from a slightly different inflationary background. The relevant information lies in the correlated pattern across $n_s$\,, $\alpha_s$\,, $r$\,, and, if it becomes accessible, $n_T$\,. Breaking this degeneracy will require either a specified inflationary potential or a joint analysis of several observables.

The slow-roll hierarchy of the new terms follows from their behaviour in the de Sitter limit. The explicit contribution to the scalar mode equation vanishes in exact de Sitter spacetime and begins at second order in the Hubble-flow functions. The scalar amplitude contains terms of order $\delta\epsilon$\,, whereas the first explicit correction to the tilt is of order $\delta\epsilon^2$ and that to the running is of order $\delta\epsilon^3$\,. No explicit $\delta$ term enters the running of the running at $\mathrm{N}^3\mathrm{LO}$\,. Thus the scale-dependent signatures are more strongly suppressed than the change in the amplitude, and the first correction to the running becomes visible only at third order.

Free tensor modes respond differently. In the absence of anisotropic stress, they obey the same propagation equation as in general relativity, but on the modified inflationary background. Scalar modes feel the same background change and the additional replacement $z_{\rm GR}\to z_{\rm eff}$\,. This difference changes both the tensor-to-scalar ratio and the single-field consistency relation. Since $r=-8n_T$ is itself only a leading-order relation in general relativity, the quantity $\Delta_c=r+8n_T$ already contains ordinary higher-order slow-roll terms; the present model adds contributions proportional to $\delta$\,. A joint measurement of $r$ and $n_T$ would therefore probe the relative scalar--tensor response to the modified dynamics \cite{Smith_2006,Boyle_2015}. Such a test remains difficult: current CMB polarisation measurements provide only an upper bound on $r$\,, and determining $n_T$ would require a detectable tensor signal over a broad range of scales \cite{BICEPKeck_2024}. LiteBIRD and the expanded Simons Observatory should improve sensitivity to primordial tensor modes, although testing the consistency relation will remain considerably harder than measuring $r$ alone \cite{LiteBIRD_2024,SimonsObservatory_2025}.

The absence of an action is the main theoretical limitation of the calculation. The modified field equations determine the time dependence of $z_{\rm eff}$\,, but not its overall multiplicative constant. We fix this freedom through the pivot prescription in Eq.~\eqref{eq:zeff_pivot_normalization} and impose the usual Bunch--Davies condition on $v_k$\,. For tensors, we separately assume the standard general-relativistic canonical normalisation and Bunch--Davies initial condition. These choices give definite amplitudes within the effective mode-equation description, but they do not replace a canonical derivation. The scalar tilt and its runnings are invariant under a constant rescaling of $z_{\rm eff}$; the absolute scalar amplitude, $r$\,, and the consistency relation are conditional on the relative scalar and tensor normalisations. A quadratic action would fix these normalisations from first principles, while a cubic action would be required for controlled predictions of primordial non-Gaussianity \cite{1992PhR...215..203M,Maldacena_2003}.

The perturbative scope must also be stated carefully. We retain the leading explicit correction in the perturbation sector, linear in $\delta\epsilon_1$\,, and expand through third order in the Hubble-flow functions. The background quantities $H$ and $\epsilon_i$\,, however, are evaluated on solutions of the modified cosmological equations without expanding those solutions in $\alpha$\,. The calculation is therefore partially resummed rather than a uniform first-order expansion in $\alpha$\,. It is valid while $|\delta\epsilon_1|\ll1$ for the relevant modes and while the background evolves smoothly. A strictly order-by-order expansion in $\alpha$ would additionally require choosing an inflationary potential and expanding the background about its general-relativistic solution.

We have calculated only the primordial spectra. A comparison with CMB temperature and polarisation data requires evolving the perturbations through reheating, radiation domination, and matter domination. Standard $\Lambda$CDM transfer functions are obtained from the general-relativistic Einstein--Boltzmann system and cannot be used unchanged if the $\alpha$-dependent terms remain relevant after inflation \cite{MaBertschinger_1995,SeljakZaldarriaga_1996,LewisChallinorLasenby_2000,classI}. The modified background and perturbation equations must instead be evolved together.

A practical next step is to choose an explicit inflationary potential and follow the modified background from horizon exit to the end of inflation. Plateau models, including Starobinsky inflation, provide useful first examples \cite{Starobinsky:1980te,Starobinsky:1979ty}. One can then determine how the correction changes the inflationary trajectory, the end of inflation, and the relation between the pivot scale and the number of e-folds. The perturbations must subsequently be evolved through reheating and the later cosmological eras and incorporated into a Boltzmann solver \cite{classI,classIII}. Only this complete evolution can provide reliable constraints on $\alpha$ and separate its effects from those of the inflationary potential and reheating history.

\appendix
\appendix

\section{Details of the
\texorpdfstring{$\mathrm{N}^3\mathrm{LO}$}{N3LO}
generalised slow-roll calculation}
\label{app:GSR_details}

This appendix gives the derivation of the scalar power spectrum presented in Section~\ref{sec:N3LO_power_spectra}. The calculation follows the generalised slow-roll construction of Refs.~\cite{Stewart_2001,Auclair_2022,Bianchi_2024}. The quantum-gravity correction enters the calculation in two ways. First, it modifies the inflationary background and hence the values and evolution of $H$\,, $\bar\varphi'$\,, and the Hubble-flow functions. Second, it changes the scalar perturbation equation through the replacement $z_{\rm GR}\to z_{\rm eff}$\,. 

The derivation proceeds in three steps. We first express the local generalised slow-roll source coefficients in terms of the corrected background quantities. These coefficients are naturally evaluated at the mode-dependent time $x_\flat=1$\,. We then re-express them at the fixed pivot time $\eta_\diamond$\,. Finally, we combine the Green's-function result with the background factor containing $z_{\rm eff}^{-2}$\,.

We work to third order in the Hubble-flow functions and retain the leading explicit perturbation-sector correction. We use
\begin{align}
    \delta(\eta)   &\coloneqq     \alpha\kappa H^2(\eta)~, & \delta_\flat &\coloneqq \alpha\kappa H_\flat^2~, & \delta_\diamond &\coloneqq \alpha\kappa H_\diamond^2~.
    \label{eq:app_delta_definitions}
\end{align}
Although the coefficients below are written in terms of $\delta$, every explicit correction contains at least one Hubble-flow function. The expansion is controlled by $|\delta\epsilon_1|\ll1$, not by $|\delta|\ll1$ alone. The background quantities themselves are evaluated on the unexpanded modified solution, as explained in Section~\ref{sec:N3LO_power_spectra}. The notation $\mathcal O(\epsilon^n)$ refers to the total degree in the Hubble-flow functions.

\subsection{Local source coefficients}
\label{app:gcoefficients}

The generalised slow-roll calculation requires the source $\eta^2z_{\rm eff}''/z_{\rm eff}-2$\,. We therefore begin by expressing $z_{\rm eff}$ and its derivatives in terms of the Hubble-flow
functions.

The normalisation condition \eqref{eq:zeff_pivot_normalization} fixes the effective background function as
\begin{align}
    z_{\rm eff}(\eta)  &=   z_{\rm GR}(\eta) \exp\left[ -\frac12 \int_{\eta_\diamond}^{\eta} \Gamma_{\rm ent}(\tilde\eta)\,  \dd\tilde\eta\right]~,  &  z_{\rm GR} &=  \frac{a\bar\varphi'}{\mathcal H}~.   \label{eq:app_zeff}
\end{align}
The lower limit ensures that
$z_{\rm eff}(\eta_\diamond)=z_{\rm GR}(\eta_\diamond)$\,. It will matter when we compute the power spectrum, but it does not enter the generalised slow-roll source because the latter depends only on derivatives of $z_{\rm eff}$\,.

The correction to the friction term in the equation for $\mathcal R_{\bm k}$ is
\begin{align}
    \Gamma_{\rm ent}(\eta)= \frac{    \left(  \mathcal H^2-\mathcal H'   \right)  \left( \sqrt{a^2+2\alpha\kappa^2\bar\varphi'^2}   - a  \right) }{ \mathcal H    \sqrt{a^2+2\alpha\kappa^2\bar\varphi'^2}  }~.
    \label{eq:app_Gamma_ent}
\end{align}
Before inserting this expression into the source, we simplify it using the corrected background relation \eqref{eq:inflation_background_identity}, which gives
\begin{align}
    \bar\varphi'^2 &=  \frac{2}{\kappa} \epsilon_1\mathcal H^2  \left( 1+\delta\epsilon_1 \right)~, & \mathcal H^2-\mathcal H' &=  \epsilon_1\mathcal H^2~.
    \label{eq:app_background_relations}
\end{align}
On the branch continuous as $\alpha\to0$, and provided $1+2\delta\epsilon_1>0$, the square-root factor therefore obeys the exact relation
\begin{align}
    \frac{  \sqrt{a^2+2\alpha\kappa^2\bar\varphi'^2}-a
    }{ \sqrt{a^2+2\alpha\kappa^2\bar\varphi'^2}  }
    =\frac{2\delta\epsilon_1}{1+2\delta\epsilon_1}~.
    \label{eq:app_square_root_expansion}
\end{align}
Substituting Eqs.~\eqref{eq:app_background_relations} and \eqref{eq:app_square_root_expansion} into Eq.~\eqref{eq:app_Gamma_ent} gives
\begin{align}
    \Gamma_{\rm ent}
    =\frac{2\delta\epsilon_1^2\mathcal H}{1+2\delta\epsilon_1}
    =2\delta\epsilon_1^2\mathcal H
    +\mathcal O\!\left[\epsilon_1\mathcal H(\delta\epsilon_1)^2\right]~.
    \label{eq:app_Gamma_expanded}
\end{align}
This expression already makes clear two things: the explicit correction begins at second order in the Hubble-flow functions and vanishes in the exact de Sitter limit.

We now cast the Mukhanov--Sasaki equation into the form used in the Green's-function expansion. Define
\begin{align}
    x  &\coloneqq  -k\eta~,
    &
    y
    &\coloneqq
    \sqrt{2k}\,v_k,
    &
    D
    &\coloneqq
    \frac{\dd}{\dd\ln x}
    =
    \eta\frac{\dd}{\dd\eta}~,
    \label{eq:app_x_y_D_def}
\end{align}
where the final equality holds at fixed $k$\,. The mode equation then becomes
\begin{align}
    \frac{\dd^2y}{\dd x^2}
    +
    \left(
        1-\frac{2}{x^2}
    \right)y
    =
    \frac{g(\ln x)}{x^2}y~,
    \qquad
    g(\ln x)
    \coloneqq
    \eta^2\frac{z_{\rm eff}''}{z_{\rm eff}}-2~.
    \label{eq:app_y_equation}
\end{align}
The homogeneous equation obtained by setting $g=0$ is the exact de Sitter equation; all departures from it are collected in the source $g$\,.

To calculate this source, it is useful to write it in terms of the logarithmic derivative of $z_{\rm eff}$\,. Since
\begin{align}
    \eta^2\frac{z_{\rm eff}''}{z_{\rm eff}}
    =
    D^2\ln z_{\rm eff}
    -
    D\ln z_{\rm eff}
    +
    \left(
        D\ln z_{\rm eff}
    \right)^2~,
\end{align}
we have
\begin{align}
    g(\ln x)
    =
    D^2\ln z_{\rm eff}
    -
    D\ln z_{\rm eff}
    +
    \left(
        D\ln z_{\rm eff}
    \right)^2
    -
    2~.
    \label{eq:app_g_operator}
\end{align}
The remaining task is therefore to find $D\ln z_{\rm eff}$ in terms of the Hubble-flow hierarchy.

From Eq.~\eqref{eq:zeff_prime_over_zeff},
\begin{align}
    \frac{z_{\rm eff}'}{z_{\rm eff}}
    =
    \mathcal H
    -
    \frac{\mathcal H'}{\mathcal H}
    +
    \frac{\bar\varphi''}{\bar\varphi'}
    -
    \frac12\Gamma_{\rm ent}~.
    \label{eq:app_zeff_prime_over_zeff_start}
\end{align}
We first evaluate the scalar-field term. Taking the logarithmic derivative of Eq.~\eqref{eq:app_background_relations} with respect to $N=\ln a$ gives
\begin{align}
    \frac{\bar\varphi''}{\bar\varphi'}
    &=
    \mathcal H
    \left(
        1-\epsilon_1+\frac12\epsilon_2
    \right)
    +
    \frac{\mathcal H}{2}
    \frac{\dd}{\dd N}
    \left(
        \delta\epsilon_1
    \right)
    +
    \mathcal H\,\mathcal O\!\left[(\delta\epsilon_1)^2\right]~.
    \label{eq:app_phipp_intermediate}
\end{align}
The required derivative follows from
\begin{align}
    \frac{\dd\epsilon_i}{\dd N}
    &=
    \epsilon_i\epsilon_{i+1}~,
    &
    \frac{\dd\delta}{\dd N}
    &=
    -2\delta\epsilon_1~,
    &
    \frac{\dd(\delta\epsilon_1)}{\dd N}
    &=
    \delta\epsilon_1
    \left(
        \epsilon_2-2\epsilon_1
    \right)~.
    \label{eq:app_N_flow_relations}
\end{align}
Hence
\begin{align}
    \frac{\bar\varphi''}{\bar\varphi'}
    =
    \mathcal H
    \left[
        1-\epsilon_1
        +
        \frac12\epsilon_2
        +
        \frac12\delta\epsilon_1
        \left(
            \epsilon_2-2\epsilon_1
        \right)
    \right]
    +
    \mathcal H\,\mathcal O\!\left[(\delta\epsilon_1)^2\right]~.
    \label{eq:app_phipp_over_phip}
\end{align}
Using
$\mathcal H'/\mathcal H=\mathcal H(1-\epsilon_1)$ and Eq.~\eqref{eq:app_Gamma_expanded} in Eq.~\eqref{eq:app_zeff_prime_over_zeff_start}, we find \begin{align}
    \frac{z_{\rm eff}'}{z_{\rm eff}}
    &=
    \mathcal H
    \left[
        1+\frac12\epsilon_2
        +
        \delta
        \left(
            \frac12\epsilon_1\epsilon_2
            -
            2\epsilon_1^2
        \right)
    \right]
    +
    \mathcal H\,\mathcal O\!\left[(\delta\epsilon_1)^2\right]~,
    \nonumber
    \\
    D\ln z_{\rm eff}
    &=
    \eta\mathcal H
    \left[
        1+\frac12\epsilon_2
        +
        \delta
        \left(
            \frac12\epsilon_1\epsilon_2
            -
            2\epsilon_1^2
        \right)
    \right]
    +
    \mathcal O\!\left[(\delta\epsilon_1)^2\right]~.
    \label{eq:app_Dlnzeff}
\end{align}
Repeated derivatives of this expression can be obtained using
\begin{align}
    D\epsilon_i
    &=
    \eta\mathcal H\,
    \epsilon_i\epsilon_{i+1}~,
    &
    D\delta
    &=
    -2\eta\mathcal H\,
    \delta\epsilon_1~.
    \label{eq:app_D_flow_relations}
\end{align}
We also require the conformal-time expansion of $\eta\mathcal H$\,. Iterating this relation
\begin{align}
    \frac{\dd}{\dd\eta}
    \left(
        \frac{1}{\mathcal H}
    \right)
    =
    -(1-\epsilon_1)~,
\end{align}
on the branch for which $\eta\mathcal H\to-1$ in the de Sitter limit gives (see \cite{Auclair_2022})
\begin{align}
    \eta\mathcal H
    &=
    -1
    -\epsilon_1
    -\epsilon_1^2
    -\epsilon_1\epsilon_2
    -\epsilon_1^3
    -3\epsilon_1^2\epsilon_2
    -\epsilon_1\epsilon_2^2
    -\epsilon_1\epsilon_2\epsilon_3
    +
    \mathcal O(\epsilon^4)~.
    \label{eq:app_etaH}
\end{align}
Equations~\eqref{eq:app_Dlnzeff}--\eqref{eq:app_etaH} now contain everything needed to evaluate Eq.~\eqref{eq:app_g_operator}. The result is a slowly varying function of $\ln x$\,. The Green's-function solution is most conveniently organised by expanding this function around $x=1$ for each mode. We denote the corresponding time by $\eta_\flat$:
\begin{align}
    x_\flat
    =
    -k\eta_\flat
    =
    1~.
    \label{eq:app_flat_time_def}
\end{align}
Since $\ln x_\flat=0$\,, the local expansion takes the form
\begin{align}
    g(\ln x)
    =
    g_{1\flat}
    +
    g_{2\flat}\ln x
    +
    g_{3\flat}\ln^2x
    +
    \mathcal O(\epsilon^4)~,
    \label{eq:app_g_local}
\end{align}
where
\begin{align}
    g_{1\flat}
    &=
    \left.g\right|_\flat~,
    &
    g_{2\flat}
    &=
    \left.Dg\right|_\flat~,
    &
    g_{3\flat}
    &=
    \frac12\left.D^2g\right|_\flat~.
    \label{eq:app_gn_taylor_def}
\end{align}
Equivalently,
\begin{align}
    g_{n\flat}
    =
    \frac{1}{(n-1)!}
    \left.
    D^{n-1}g
    \right|_\flat~.
\end{align}
Carrying out these derivatives through third order gives Eqs.~\eqref{eq:g1_main}--\eqref{eq:g3_main}.

Setting $\alpha=0$ reproduces the standard single-field source coefficients \cite{Auclair_2022}. In addition, the explicit quantum-gravity terms appear in $g_{1\flat}$ and $g_{2\flat}$\,, but not in $g_{3\flat}$ at the order retained. This follows directly from Eq.~\eqref{eq:app_Gamma_expanded}: the correction already starts at second order in the Hubble-flow functions, and each derivative raises the slow-roll order by one.

\subsection{Transport from the mode-dependent time to the pivot}
\label{app:pivot_change}

The local expansion above is carried out at $x_\flat=1$\,, which is a different time for each Fourier mode. This is the natural choice for the Green's-function calculation, but it is not suitable for comparing with the observations. The final spectrum must instead be written in terms of background quantities evaluated at a single pivot time.

We define $\eta_\diamond$ by
\begin{align}
    -k_\diamond\eta_\diamond
    =
    1~.
    \label{eq:app_pivot_time_def}
\end{align}
Since $\eta_\flat=-1/k$ and $\eta_\diamond=-1/k_\diamond$\,, the logarithmic distance from the pivot is
\begin{align}
    L
    \coloneqq
    \ln\left(
        \frac{k}{k_\diamond}
    \right)
    =
    -\ln\left(
        \frac{\eta_\flat}{\eta_\diamond}
    \right)~.
    \label{eq:app_L_eta_relation}
\end{align}
Varying $k$ therefore moves the evaluation time $\eta_\flat$\,. Along this family of times,
\begin{align}
    \frac{\dd}{\dd L}
    =
    -\frac{\dd}{\dd\ln|\eta_\flat|}
    =
    -\eta_\flat\frac{\dd}{\dd\eta_\flat}
    =
    -\eta_\flat\mathcal H_\flat
    \frac{\dd}{\dd N_\flat}~.
    \label{eq:app_d_dL_operator}
\end{align}
This operator allows us to Taylor expand every quantity evaluated at $\eta_\flat$ around $\eta_\diamond$\,.

Applying Eq.~\eqref{eq:app_d_dL_operator} repeatedly to the
Hubble-flow functions gives
\begin{align}
\epsilon_{i\flat}
&=
\epsilon_{i\diamond}
+
\epsilon_{i\diamond}\epsilon_{i+1\diamond}
\left(
    1+\epsilon_{1\diamond}
    +\epsilon_{1\diamond}^2
    +\epsilon_{1\diamond}\epsilon_{2\diamond}
\right)L
\nonumber
\\
&\quad
+
\frac12
\epsilon_{i\diamond}\epsilon_{i+1\diamond}
\left(
    \epsilon_{i+1\diamond}
    +\epsilon_{i+2\diamond}
    +\epsilon_{1\diamond}\epsilon_{2\diamond}
    +2\epsilon_{1\diamond}\epsilon_{i+1\diamond}
    +2\epsilon_{1\diamond}\epsilon_{i+2\diamond}
\right)L^2
\nonumber
\\
&\quad
+
\frac16
\epsilon_{i\diamond}\epsilon_{i+1\diamond}
\left(
    \epsilon_{i+1\diamond}^2
    +\epsilon_{i+2\diamond}^2
    +3\epsilon_{i+1\diamond}\epsilon_{i+2\diamond}
    +\epsilon_{i+2\diamond}\epsilon_{i+3\diamond}
\right)L^3
+
\mathcal O(\epsilon^5)~.
\label{eq:app_epsilon_transport}
\end{align}
Although the final spectrum is required only through cubic order, some terms of fourth order must be retained in this transport formula because they can multiply lower-order prefactors.

The same procedure gives the Hubble-rate transport. Starting from $\dd\ln H^2/\dd N=-2\epsilon_1$\,, we obtain
\begin{align}
    \frac{H_\flat^2}{H_\diamond^2}
    &=
    1
    -
    2
    \left(
        \epsilon_{1\diamond}
        +
        \epsilon_{1\diamond}^2
        +
        \epsilon_{1\diamond}^3
        +
        \epsilon_{1\diamond}^2\epsilon_{2\diamond}
    \right)L
    +
    \left(
        2\epsilon_{1\diamond}^2
        -
        \epsilon_{1\diamond}\epsilon_{2\diamond}
        +
        4\epsilon_{1\diamond}^3
        -
        3\epsilon_{1\diamond}^2\epsilon_{2\diamond}
    \right)L^2
    \nonumber
    \\
    &\quad
    +
    \frac16
    \left(
        -8\epsilon_{1\diamond}^3
        +
        12\epsilon_{1\diamond}^2\epsilon_{2\diamond}
        -
        2\epsilon_{1\diamond}\epsilon_{2\diamond}^2
        -
        2\epsilon_{1\diamond}
        \epsilon_{2\diamond}\epsilon_{3\diamond}
    \right)L^3
    +
    \mathcal O(\epsilon^4)~.
    \label{eq:app_H_transport}
\end{align}
Since $\delta=\alpha\kappa H^2$\,, its transport follows from the same expression. Terms proportional to $\delta$ in the source coefficients begin at second slow-roll order, for which the leading transport correction would suffice. The background normalisation factor in Eq.~\eqref{eq:app_zflat_background_factor}, however, already contains $\delta_\flat\epsilon_{1\flat}$ at first order. To obtain that contribution consistently through cubic slow-roll order, we need
\begin{align}
    \delta_\flat
    =
    \delta_\diamond
    \left(
        1
        -2\left(\epsilon_{1\diamond}+\epsilon_{1\diamond}^2\right)L
        +\left(2\epsilon_{1\diamond}^2
        -\epsilon_{1\diamond}\epsilon_{2\diamond}\right)L^2
    \right)
    +
    \mathcal O(\delta_\diamond\epsilon^3)~.
    \label{eq:app_delta_transport}
\end{align}
Equations~\eqref{eq:app_epsilon_transport}--\eqref{eq:app_delta_transport} express all mode-dependent background quantities in terms of their values at the pivot.

\subsection{The factor
\texorpdfstring{$z_{\rm eff}^{-2}$}{zeff inverse squared}
and the final spectrum}
\label{app:entropy_normalization}

The source coefficients determine the evolution of $v_k$\,, but the observable curvature perturbation is $\mathcal R_k=v_k/z_{\rm eff}$\,. We must therefore restore the factor $z_{\rm eff}^{-2}$ when constructing the power spectrum:
\begin{align}
    \mathcal P_{\mathcal R}(k)
    &=
    \frac{k^3}{2\pi^2}
    \left|
        \frac{v_k}{z_{\rm eff}}
    \right|_{x\to0}^{2}
    =
    \frac{1}{4\pi^2}
    \lim_{x\to0^+}
    \left|
        x\,y(x)
    \right|^2
    \left|
        \frac{k}{x z_{\rm eff}}
    \right|^2~.
    \label{eq:app_PR_split}
\end{align}
The first factor follows from the Green's-function solution and depends on the background through $g_{n\flat}$\,. The second factor supplies the normalisation associated with the corrected background function. Both must be expanded about the same time $x_\flat=1$ before they are transported to the pivot.

At this time, Eq.~\eqref{eq:app_zeff} gives
\begin{align}
    z_{{\rm eff},\flat}^{-2}
    =
    \frac{\mathcal H_\flat^2}
    {a_\flat^2\bar\varphi_\flat'^2}
    \exp\left[
        \int_{\eta_\diamond}^{\eta_\flat}
        \Gamma_{\rm ent}(\eta)\,
        \dd\eta
    \right]~.
    \label{eq:app_zeff_inverse_flat}
\end{align}
Using the corrected background relation \eqref{eq:app_background_relations}, we find
\begin{align}
    \frac{\mathcal H_\flat^2}
    {a_\flat^2\bar\varphi_\flat'^2}
    &=
    \frac{\kappa}
    {2a_\flat^2\epsilon_{1\flat}}
    \frac{1}{1+\delta_\flat\epsilon_{1\flat}}
    =
    \frac{\kappa}
    {2a_\flat^2\epsilon_{1\flat}}
    \left[
        1-\delta_\flat\epsilon_{1\flat}
        +\mathcal O\!\left((\delta_\flat\epsilon_{1\flat})^2\right)
    \right]~,
    &
    \frac{k^2}{a_\flat^2}
    &=
    \frac{H_\flat^2}
    {(\eta_\flat\mathcal H_\flat)^2}~,
    \label{eq:app_background_prefactor_relations}
\end{align}
where we used $x_\flat=1$\,, or $k=-1/\eta_\flat$\,. It follows that the background prefactor entering the generalised slow-roll result can be written as
\begin{align}
    \left.
    \left|
        \frac{k}{x z_{\rm eff}}
    \right|^2
    \right|_{x_\flat=1}
    =
    \frac{\kappa H_\flat^2}
    {2\epsilon_{1\flat}}
    \frac{1}
    {(\eta_\flat\mathcal H_\flat)^2}
    \left[
        1-\delta_\flat\epsilon_{1\flat}
        +\mathcal O\!\left((\delta_\flat\epsilon_{1\flat})^2\right)
    \right]
    \exp\left[
        \int_{\eta_\diamond}^{\eta_\flat}
        \Gamma_{\rm ent}(\eta)\,
        \dd\eta
    \right]~.
    \label{eq:app_zflat_background_factor}
\end{align}
The factors involving $H_\flat$\,, $\epsilon_{1\flat}$\,, and $\eta_\flat\mathcal H_\flat$ generate both the standard slow-roll terms and the corrections induced by the modified background. The remaining exponential is the direct contribution from the replacement $z_{\rm GR}\to z_{\rm eff}$\,.

To evaluate it, define
\begin{align}
    I_\Gamma(k)
    \coloneqq
    \int_{\eta_\diamond}^{\eta_\flat}
    \Gamma_{\rm ent}(\eta)\,
    \dd\eta~.
    \label{eq:app_I_Gamma_def}
\end{align}
Using Eq.~\eqref{eq:app_Gamma_expanded} and $\dd N=\mathcal H\,\dd\eta$\,, this becomes
\begin{align}
    I_\Gamma(k)
    =
    2
    \int_{N_\diamond}^{N_\flat}
    \delta(N)\epsilon_1^2(N)\,
    \dd N
    +
    \mathcal O\!\left[(\delta\epsilon_1)^2\right]~.
    \label{eq:app_I_Gamma_N}
\end{align}
The upper limit depends on $k$\,. Differentiating it with respect to $L$ gives
\begin{align}
    \frac{\dd N_\flat}{\dd L}
    =
    -\eta_\flat\mathcal H_\flat~.
    \label{eq:app_dNflat_dL}
\end{align}
We use this relation together with the flow equations to expand the integral around $L=0$\,. Through third order,
\begin{align}
    I_\Gamma(k)
    &=
    2\delta_\diamond\epsilon_{1\diamond}^2
    \left(
        1+\epsilon_{1\diamond}
    \right)L
    +
    2\delta_\diamond\epsilon_{1\diamond}^2
    \left(
        \epsilon_{2\diamond}
        -
        \epsilon_{1\diamond}
    \right)L^2
    +
    \mathcal O\!\left(\delta_\diamond\epsilon^4,
    (\delta_\diamond\epsilon_{1\diamond})^2\right)~.
    \label{eq:app_entropy_exponent_L}
\end{align}
Because the retained contribution to $I_\Gamma$ is linear in $\delta$ and beginning at order $\delta_\diamond\epsilon^2$\,, its square lies beyond the accuracy of the calculation. We may therefore expand
\begin{align}
    \exp\left[
        I_\Gamma(k)
    \right]
    &=
    1
    +
    2\delta_\diamond\epsilon_{1\diamond}^2
    \left(
        1+\epsilon_{1\diamond}
    \right)L
    +
    2\delta_\diamond\epsilon_{1\diamond}^2
    \left(
        \epsilon_{2\diamond}
        -
        \epsilon_{1\diamond}
    \right)L^2
    +
    \mathcal O\!\left(\delta_\diamond\epsilon^4,
    (\delta_\diamond\epsilon_{1\diamond})^2\right)~.
    \label{eq:app_entropy_expansion_L}
\end{align}
At the pivot, $L=0$ and this factor is unity, as required by $z_{\rm eff}(\eta_\diamond)=z_{\rm GR}(\eta_\diamond)$\,.

We can now assemble the result. The Green's function integrals supply universal numerical constants. The corrected source coefficients $g_{n\flat}$ determine the mode evolution. The transport formulae replace all quantities evaluated at $\eta_\flat$ by pivot quantities, while Eq.~\eqref{eq:app_zflat_background_factor} supplies the background normalisation. Expanding their product consistently through $\mathrm{N}^3\mathrm{LO}$ gives
\begin{align}
    \mathcal P_{\mathcal R}(k)
    =
    \frac{\kappa H_\diamond^2}
    {8\pi^2\epsilon_{1\diamond}}
    \left[
        P_0
        +
        P_1L
        +
        P_2L^2
        +
        P_3L^3
        +
        \mathcal O\!\left(\epsilon^4,(\delta_\diamond\epsilon_{1\diamond})^2\right)
    \right]
    ~,
    \qquad
    L
    =
    \ln\left(
        \frac{k}{k_\diamond}
    \right)~.
    \label{eq:app_PR_pivot}
\end{align}
This is Eq.~\eqref{eq:PR_main}. The coefficients $P_0$\,, $P_1$\,, $P_2$\,, and $P_3$ are given in Eqs.~\eqref{eq:P0_main}--\eqref{eq:P3_main}.

For consistency, one can see that when $\alpha=0$\,, both $\Gamma_{\rm ent}$ and $\delta$ vanish, so $z_{\rm eff}=z_{\rm GR}$ and the standard single-field $\mathrm{N}^3\mathrm{LO}$ spectrum is recovered \cite{Auclair_2022}. At $k=k_\diamond$\,, the integral $I_\Gamma$ vanishes, in agreement with the normalisation chosen for $z_{\rm eff}$\,. Finally, in the exact de Sitter limit, $\epsilon_1=0$\,, the explicit quantum-gravity correction disappears from both the source and the background prefactor.

\acknowledgments

VK thanks Oliver Friedrich for insightful discussions. VK gratefully acknowledges support from the \emph{Deutscher Akademischer Austauschdienst} (DAAD, German Academic Exchange Service) and from a Short-Term Scientific Mission (STSM) grant awarded through COST Action CA23115, \emph{Relativistic Quantum Information} (RQI), supported by COST (European Cooperation in Science and Technology). The STSM supported a research visit during which part of this work was developed. MdC acknowledges support from the INFN Iniziativa Specifica GeoSymQFT, QUAGRAP, and QGSKY. This work also contributes to COST Action CA23130, \emph{Bridging high and low energies in search of quantum gravity} (BridgeQG).

\bibliography{bibliography_corrected}
\bibliographystyle{unsrtnat}
\end{document}